\documentclass[twocolumn,superscriptaddress,floatfix,prx,showpacs,nolongbibliography]{revtex4-2}
\usepackage{graphicx}
\usepackage{amsmath}
\usepackage{amssymb}
\usepackage{hyperref}
\usepackage{color}
\usepackage{braket}
\usepackage{dsfont}
\usepackage{lineno}
\usepackage{orcidlink}
\usepackage{float}

\newcommand{\nint}{\operatorname{nint}}
\begin{document}
\title{A Clue on Small-Capacitance Josephson Junction: What to Expect from Cooper Pair Ideal Conductor and Ohmic Resistor in Parallel?}

\author{Francesco Giuseppe Capone\orcidlink{0009-0000-9832-4385}}
\affiliation{Dip. di Fisica E. Pancini - Università di Napoli Federico II - I-80126 Napoli, Italy}
\affiliation{INFN, Sezione di Napoli - Complesso Universitario di Monte S. Angelo - I-80126 Napoli, Italy}

\author{Antonio de Candia\orcidlink{0000-0002-9869-1297}}
\affiliation{INFN, Sezione di Napoli - Complesso Universitario di Monte S. Angelo - I-80126 Napoli, Italy}
\affiliation{SPIN-CNR and Dip. di Fisica E. Pancini - Università di Napoli Federico II - I-80126 Napoli, Italy}
\author{Vittorio Cataudella\orcidlink{0000-0002-1835-1429}}
\affiliation{INFN, Sezione di Napoli - Complesso Universitario di Monte S. Angelo - I-80126 Napoli, Italy}
\affiliation{SPIN-CNR and Dip. di Fisica E. Pancini - Università di Napoli Federico II - I-80126 Napoli, Italy}
\author{Naoto Nagaosa\orcidlink{0000-0001-7924-6000}}
\affiliation{RIKEN Center for Emergent Matter Science (CEMS), Wako, Saitama 351-0198, Japan}
\affiliation{Fundamental Quantum Science Program, TRIP Headquarters, RIKEN, Wako 351-0198, Japan}
\author{Carmine Antonio Perroni\orcidlink{0000-0002-3316-6782}}
\affiliation{INFN, Sezione di Napoli - Complesso Universitario di Monte S. Angelo - I-80126 Napoli, Italy}
\affiliation{SPIN-CNR and Dip. di Fisica E. Pancini - Università di Napoli Federico II - I-80126 Napoli, Italy}
\author{Giulio De Filippis\orcidlink{0000-0003-0557-3556}}
\affiliation{INFN, Sezione di Napoli - Complesso Universitario di Monte S. Angelo - I-80126 Napoli, Italy}
\affiliation{SPIN-CNR and Dip. di Fisica E. Pancini - Università di Napoli Federico II - I-80126 Napoli, Italy}

\begin{abstract}
By using analytical and Worldline Monte Carlo approaches, we investigate the effects induced by quantum phase fluctuations combined with quasiparticle subgap and shunt resistances on a small-capacitance Josephson junction. By using the linear response theory in the presence of two biasing schemes, we prove that the ideal conduction, foreseen in the pioneering papers on this topic, is not robust against either quantum phase fluctuations or dissipative effects. By including both of them in the Hamiltonian, we prove that an increase of the Ohmic dissipation strength induces a Berezinskii-Kosterlitz-Thouless quantum phase transition in thermodynamic equilibrium. Then we study charge and phase fluctuations at the thermodynamic equilibrium within the linear response theory. We find that the phase particle motion, in a quantum Josephson junction, does not change from diffusive to localized, resulting in an insulator-superconductor transition, as is commonly believed. At the transition, we prove that: i) the motion of the phase particle changes from ballistic to localized; ii) by turning on the coupling with the environment, a long-lived excitation at finite frequency emerges in the charge response function: it evolves first into a resonance and then disappears at the transition. Consequences beyond the linear response regime are investigated, leading to an alternative comprehensive physical picture for this system: we predict a transition from a dissipative quasiparticle current to a polaronic Cooper pair current. 

\end{abstract}
\maketitle

\section{Introduction}
The study of open quantum systems is crucial in many research fields, ranging from condensed matter theory and quantum transport~\cite{Weiss1999, nitzan_chemical, Leggett_spinboson} to quantum chemistry~\cite{may_molecular_systems, nitzan_chemical}, quantum information~\cite{nielsen_qc_qinfo} and quantum metrology~\cite{Alipour_metrology}. In general, the interaction of an open quantum system with the degrees of freedom of the environment induces decoherence, dissipation, relaxation towards the thermodynamic equilibrium states, i.e. a loss of the characteristic quantum features~\cite{gardiner_zoller_QuantumNoise, Lidar}. However, there are cases where memory effects play a key role, allowing the study of important quantum features, such as coherence, resonances, and entanglement~\cite{Breuer_RevModPhys_nonmarkovian, Rivas_nonmarkvovian, Maniscalco_nonmarkovian}. The interaction with the environment can also induce equilibrium and dynamical quantum phase transitions. The well-known spin-boson~\cite{Leggett_spinboson, Weiss1999} and quantum Rabi~\cite{Rabi_original, JCM_rabi} models have attracted considerable interest during recent decades because of their simple experimental setups and relevance in different research fields, in particular within quantum optics~\cite{Rabi_optics1, Rabi_optics2, Rabi_optics3}. In these cases, characterized by a two-level system, the existence of dissipation-driven quantum phase transitions has been extensively clarified~\cite{Giulio_spinboson_PRB, Giulio_manyspin_PRB, Grazia_NatComm, Giulio_PRL}. On the other hand, the occurrence of a quantum phase transition in a Josephson junction and its capacitor, analogous to a massive particle in a periodic potential, coupled to a bath of harmonic oscillators, has caused a long-standing controversy~\cite{Murani_absence, Comment_on_absence, Reply_to_comment}. In fact, in pioneering papers on this subject,  the so-called dissipative phase transition has been predicted~\cite{Schmid, Bulgadaev}. 

The underlying principle is the behavior of macroscopic quantum tunneling of the phase as a function of the strength of the interaction with a dissipative quantum-mechanical environment, described by one or two bosonic fields: at weak coupling, macroscopic quantum tunneling takes place, destroying superconductivity of a junction, whereas suppression of tunneling, occurring at high couplings with the environment, restores the Josephson current. Different experimental attempts~\cite{Exp_1, Exp_2, Exp_3} have been made to observe the theoretically predicted dissipation-driven phase transition in a small Josephson junction. On the other hand, interpretation of these results is still debated. Indeed, recently, even the absence of this quantum phase transition in the predicted parameter regime has been reported~\cite{Murani_absence}. Despite many years of research, an outright understanding of this quantum phase transition, driven by the coupling with environmental degrees of freedom, has yet to be achieved. 

In pioneering works on this issue~\cite{Josephson1, Josephson2, Likharev, Ambegaokar_Josephson_effect}, the phase variable of a single Josephson junction has been treated as a classical variable. The well-known Josephson equations, $I(t)=I_c(t)\sin(\varphi(t))$ and $\frac{\partial \varphi(t)}{\partial t}=\frac{2e}{\hbar}V(t)$, $I(t)$ and $V(t)$ being the current and the voltage through the junction, admit, for $I<I_c$ ($I_c$ is the critical current), a stationary solution with $\varphi$ independent of $t$, corresponding to the superconducting state. Taking into account the quantum phase fluctuations and the effects of the interaction with the dissipative environment was the next step. To this end, two different Hamiltonians have been proposed in the literature~\cite{Schon_Zaikin_review, Fazio_review, PhysRevB_Schon, PhysRevLett_Schon}. The main prediction is that, by varying the strength of the coupling with the environmental degrees of freedom, an insulator-superconductor transition takes place, where the motion of the phase particle changes from diffusive to localized\cite{Schmid, Bulgadaev}.

Here, first of all, we discuss the two models known in the literature (Section II). Later (Section III), by using the linear response theory and taking into account two biasing schemes, we derive two Kubo-Mori formulas relating current and voltage across the junction. In Sections IV and V, by using analytical and numerically exact approaches, we prove that the ideal conduction, foreseen in the pioneering papers on this topic, is not robust against either quantum phase fluctuations or dissipative effects. In section VI, by including both of them in the Hamiltonian and using worldline Monte Carlo approaches, we find that an increase of the Ohmic dissipation strength induces localization of the phase particle, resulting in a Berezinskii-Kosterlitz-Thouless (BKT) quantum phase transition at the thermodynamic equilibrium in both the considered models. In Section VII, we investigate phase and charge fluctuations at the thermodynamic equilibrium within the linear response theory through a current bias. We prove that differently from the previous predictions: i) at the transition the particle phase motion changes from ballistic to localized; ii) a gap opens in the charge and current response function spectra for any value of the coupling with the environment; iii) by turning on the interaction with the bosonic degrees of freedom, a long-lived excitation emerges at finite frequency: it evolves first into a resonance and then disappears at the transition. Finally, we discuss consequences of all these findings beyond the linear response regime: we predict a transition from dissipative quasiparticle current to polaronic Cooper pair current. 

\section{The model}
The Josephson effect takes place in the presence of a weak electrical contact of two superconducting samples, i.e. a thin layer of a non-superconducting material between two layers of superconducting material.  It is a typical example of macroscopic quantum phenomenon.  Within pioneering works~\cite{Josephson1, Josephson2, Feynman3}, this system has been represented as a nonlinear inductor with phase difference $\varphi$ between superconductors treated as a classical variable.  Later, the effects that arise from phase fluctuations across the junction, quasiparticle subgap, and shunt resistances in Josephson tunneling have been taken into account, leading to two different models. 

A microscopic model, accounting for phase fluctuations and quasiparticle (QP) tunneling due to the existence of a subgap resistance $R_{QP}$ \cite{PhysRevLett_Schon, PhysRevB_Schon}, leads to an effective action for the phase difference between the two superconductors:
\begin{eqnarray}\label{eqn:action_QP}
&&\mathcal{S}_{QP}[\varphi(\tau)]=\frac{\hbar^2}{4E_C}\int_0^{\beta \hbar} d \tau \, \biggl(\frac{d\varphi}{d\tau}\biggr)^2 -E_J\int_0^{\beta \hbar} d \tau \, \cos\varphi(\tau) \nonumber \\ && + \int_0^{\beta \hbar} d \tau  \int_0^{\beta \hbar} d\tau' \, K(\tau-\tau')\sin^2\biggl( \frac{\varphi(\tau)-\varphi(\tau')}{4}\biggr).
\end{eqnarray}
Here $E_C$ describes the capacitive Coulomb interaction of charges accumulating in the vicinity of the oxide barrier (it depends on the geometry and properties of the insulator), $E_J$ is the Josephson coupling energy ($E_J=\frac{\hbar I_c}{2e}$) and the kernel $K(\tau)$, representing a retarded potential, is given by: 
\begin{equation*}
    K(\tau) = \bigl(2\hbar/\pi^2\bigr)\int_0^{+\infty}d\omega \,J(\omega) D_\omega (\tau).  
\end{equation*}
The spectral function $J(\omega)$ encompasses the whole physics of the system and the function 
\begin{equation*}
D_\omega(\tau) = \cosh\bigl[\omega(\beta\hbar/2 -|\tau|)\bigr]/\sinh(\beta\hbar\omega/2) 
\end{equation*}
represents the propagator of a harmonic oscillator with frequency $\omega$. We emphasize that the trigonometric dependence on the phase difference in the action reflects the discreteness of charge that tunnels across the thin insulating barrier: it describes the tunneling of single electrons. 

The presence of Ohmic shunt resistance $R_S$ is taken into account in a phenomenological model {a l\'a} Caldeira-Leggett (CL) \cite{PhysRevLettCL1}. In this case, the trigonometric function in the action is replaced by its quadratic expansion, allowing a continuous change of the charge. In both models, $J(\omega)$  depends linearly on frequency up to a cut-off frequency $\omega_D$: $J(\omega) = \alpha\,\omega \Theta(\omega_D-\omega) $, $\alpha$ being the ratio between the quantum resistance $h/4e^2$ and the intrinsic subgap resistance $R_{QP}$ ($R_S$) in the QP (CL) model: 

\begin{equation*}
\alpha = \frac{h}{4 e^2 R_I}, I=QP,S.
\end{equation*}

It can be shown that QP action and CL action can be derived respectively from the following Hamiltonians:
\begin{eqnarray}
            &&H_{QP} = \frac{Q^2}{2C} - E_J\cos\biggl(\frac{\phi}{\phi_0}2\pi\biggl) + \sum_{i=1}^N\biggl[\frac{q^2_{i,1}}{2C_{i,1}}+\frac{\phi^2_{i,1}}{2L_{i,1}}\biggr] \nonumber \\&& + \sum_{i=1}^N\biggl[\frac{q^2_{i,2}}{2C_{i,2}}+\frac{\phi^2_{i,2}}{2L_{i,2}}\biggr]  + \sum_{i=1}^N \cos\biggl(\frac{\phi}{\phi_0}\frac{2\pi}{2}\biggr)S_{i,1}\phi_{i,1} \nonumber\\
             &&+  \sum_{i=1}^N \sin\biggl(\frac{\phi}{\phi_0}\frac{2\pi}{2}\biggr)S_{i,2}\phi_{i,2},
\end{eqnarray}
\begin{equation}
            H_{CL} = \frac{Q^2}{2C} - E_J\cos\biggl(\frac{\phi}{\phi_0}2 \pi\biggr)  + \sum_{i=1}^N \biggl[ \frac{q_i^2}{2C_i} + \frac{1}{2L_i}\bigl(\phi_i -\phi\bigr)^2\biggr],
\end{equation}
\begin{figure}
\flushleft
        \includegraphics[scale=0.50]{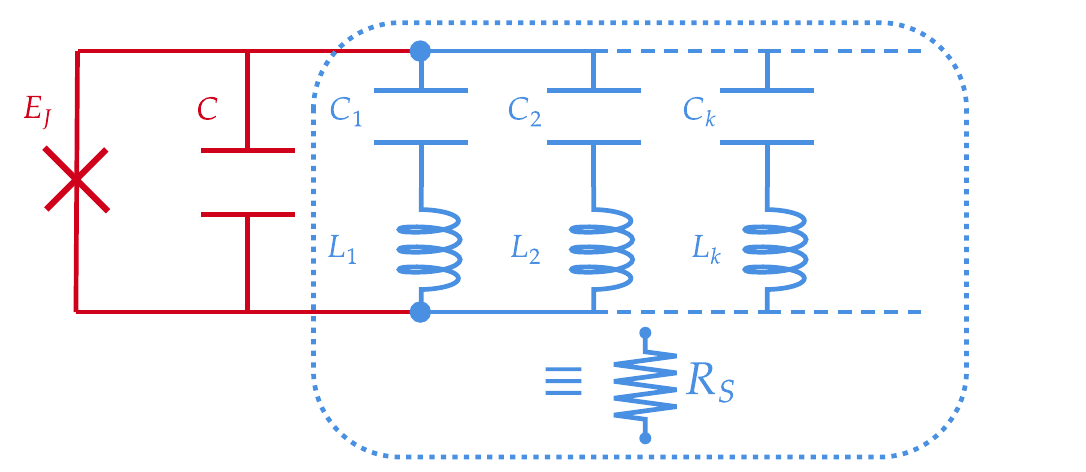}
        \caption{Sketch of the model {a l\'a} Caldeira-Leggett: circuit consisting of a nonlinear inductor $E_J$, a capacitor $C$ (the quantum junction depicted in red), and a shunt resistance $R_S$ in parallel (depicted in blue). Here, $R_S$ is represented as an infinite set of LC oscillators. In contrast, such representation is not possible for the subgap resistance $R_{QP}$ in the QP model, where the discrete charge flow is instead described by a nonlinear coupling between the nonlinear inductor and two bosonic baths.} 
\label{fig:0}
\end{figure}
where $C = 2e^2/E_C$ represents the capacitance of the junction. The flux $\phi = \frac{\phi_0}{2\pi}\varphi$  ($\phi_0=h/2e$ is the magnetic flux quantum) and the charge $Q$ on the junction electrodes are conjugate variables, i.e. $[\phi,Q]=i\hbar$. Dissipation is included in these Hamiltonians through coupling of the junction's degrees of freedom to a single bosonic bath (CL) or two bosonic baths (QP), with coupling strengths given by $S_{i,j}$ in the latter case. We emphasize that $\phi_{i,j}$ and $q_{i,j}$ are conjugate variables in these ensembles of charged harmonic oscillators, each of them describing a LC circuit. In the mechanic analog, the flux plays the role of the position coordinate and the charge the role of the momentum coordinate.  The spectral function of the QP model is defined as $\frac{2\hbar^2}{\pi^2}J(\omega) = \sum_{i=1}^NS_{i,j}^2\frac{\hbar}{2C_{i,j}\omega_{i,j}}\delta(\omega-\omega_{i,j})$ where $\omega^2_{i,j} = \frac{1}{L_{i,j}C_{i,j}}$ for $j=1,2$ (note that $J(\omega)$ is independent of $j$). Equivalently, for the CL model, we define $J(\omega) = \frac{\pi}{2} \sum_{i=1}^N  \frac{\omega_i}{L_i}\frac{h}{4e^2}\delta(\omega-\omega_i)$ where $\omega^2_i = \frac{1}{L_iC_i}$. These equations point out the physical meaning of the dimensionless parameter $\alpha$: it measures the strength of the interaction with the environment. 

The classical motion equation for the variable $Q$, in the presence of current bias, proves that the CL and QP models describe a parallel of a nonlinear inductor, a capacitor and a resistor. In particular, within the QP model, the junction-resistor coupling is non-linear. We also emphasize that the cutoff frequency $\omega_D$ is associated with the highest energy scale in the problem. Within the QP model, $\omega_D$ is related to the Fermi energy of the two superconductors. On the other hand, within the CL model it ensures that the conductance of the shunt resistance, given by a set of infinite LC oscillators in parallel\cite{devoret} (see Fig.\ref{fig:0}), is independent of the frequency up to $\omega_D$, i.e. it is governed by the Ohmic $I-V$ relation. In the following, we will fix $\hbar\omega_D=50E_C$. Finally, it is worth mentioning that the phase variable $\varphi$ must be interpreted within the "extended picture" \cite{Likharev}. Indeed, if the superconducting tunnel junction is biased by an external current, we have to add an additional potential contribution -$I\varphi$ in the Hamiltonian. This allows for distinguishing states where the phase difference is a multiple of $2\pi$. Consequently, the Hilbert space of the system must retain the same structure even in equilibrium, as the current approaches zero. 
\section{Linear Transport}
The linear response theory provides the main criterion to discriminate between insulators, metals, and superconductors\cite{Fetter}. Due to the lack, in the literature on this subject, of a precise description of transport features, in the following we will derive the two principal Kubo-Mori formulas relating voltage, $V$, and current, $I$, across the junction (more details are given in Appendix \ref{sec:app_transport}). 

First, we define $I$ and $V$ in the Heisenberg picture: $I(t) = -\dot{Q}$ and $V(t) = \dot{\phi} = Q/C$. Then we describe two biasing schemes. The first one is induced by a voltage $\delta V(t)$ across the junction for $t>0$, i.e. we add a small perturbation $I \delta\phi(t)$ to the Hamiltonian of the system where $\delta V(t) = \frac{d}{dt} \delta \phi(t)$ and $\delta \phi(t)$ is a classical function of time. We assume that for $t<0$ the system is in thermodynamic equilibrium. At first order in the perturbation, the total induced current across the junction is $ I_{\text{tot}}(t)  =  I(t)  +\frac{\partial I}{\partial \phi}\delta \phi (t) $. From linear response theory it follows that variation of the total current is \cite{Mahan}:
\begin{equation}\label{eqn:linear_current_tot}
\delta \langle I_{\text{tot}}\rangle (t)  = \int_{-\infty}^{+\infty}dt' \,\Pi_I(t-t')\delta\phi(t') + \biggl\langle\frac{\partial I}{\partial \phi}\biggr\rangle\delta\phi(t),           
\end{equation}
where $\Pi_I(t-t')$ is the current-current correlation function: $\Pi_I(t-t')=-\frac{i}{\hbar}\theta(t-t')\bigl\langle[I(t),I(t')] \bigr\rangle$, $\theta(t)$ being the Heaviside function. Defining the Fourier transform of the linear admittance as $Y(z) = \int_0^{+\infty}dt \,Y(t)e^{izt} = \delta\langle I_{\text{tot}}\rangle(z)/\delta V(z)$, from \eqref{eqn:linear_current_tot} it is straightforward to prove that:
\begin{equation}\label{eqn:kubo_current}
Y(z) = \frac{i}{z}\biggl[\Pi_I(z)+\biggl\langle\frac{\partial I}{\partial\phi}\biggl\rangle\biggr],   
\end{equation}
where $z$ lies in the complex upper half plane, i.e. $z = \omega +i\epsilon$, with $\epsilon >0$.

This equation for the admittance can be generalized, within the Kubo-Mori formalism, to any response function, $\Psi_{A}(z)$, involving a generic observable $A$ (see Appendix \ref{sec:app_transport}). In the case $A=I$, the response function is the admittance: $\Psi_{I}(z)=Y(z)$.  By taking the limit $\epsilon \to 0$ , the real part of the response function turns out to be\cite{Shastry_PRB, Evertz_PRB}:
\begin{equation}\label{eqn:kubo_tot}
\Re\bigl[\Psi_I(\omega)\bigr] = D_I\delta(\omega) + \Psi_{I,\text{reg}}(\omega).
\end{equation}
Here, even function $\Psi_{I,\text{reg}}(\omega)$ is the regular part of the conductivity and $D_I$, the strength of the delta function, is the Drude weight or charge stiffness. By introducing the Matsubara Green function associated with the observable $I$,  $\Pi_I(\tau) = -\frac{1}{\hbar} \bigl\langle I(\tau)I(0)\bigr\rangle $ ($\tau$ is imaginary time in the range $-\hbar\beta<\tau<\hbar\beta$ and $T_{\tau}$ is the time ordering operator), and the corresponding Fourier coefficients $\Pi_I(i\omega_n) = -\frac{1}{\hbar}\int_0^{\beta\hbar}d\tau \,\bigl\langle I(\tau)I(0)\bigr\rangle e^{i\omega_n \tau}$, $\omega_n$ being the Matsubara frequencies, it is possible to prove\cite{Evertz_PRB, Shastry_PRB}:

\begin{equation}\label{eqn:Drude}
    D_I = \biggl\langle \frac{\partial I}{\partial\phi}\biggr\rangle + \Pi_{I}(i\omega_n \to 0).
\end{equation}
A property closely related to the Drude weight is the Meissner stiffness, defined by: 
\begin{equation}\label{eqn:Drude1}
    D_{I,M} = \biggl\langle \frac{\partial I}{\partial\phi} \biggr\rangle + \Pi_{I}(i\omega_n =0).
\end{equation}
The difference between the two stiffnesses is a sum over all degenerate manifolds: 
\begin{equation}\label{eqn:Differenza}
    D_{I,M}-D_I = -\beta \sum_{E_n=E_m} p_n|\bigl \langle n| I | m \bigr \rangle|^2,
\end{equation}
$p_n$ being the Boltzmann weight of the eigenstate $|n\rangle$.  The values of $D_I$ and $D_{I,M}$  allow to discriminate insulators, metals and superconductors:  $D_I$ and $D_{I,M}$ are both vanishing in insulators at any temperature; $D_{I,M}=0$ and $D_{I}\ne 0$ in metals at zero temperature (at finite temperature $D_I=0$ and the regular part of the conductivity exhibits, in the limit $\omega \rightarrow 0$, a non vanishing value that decreases by increasing the temperature), and $D_{I,M}\ne0$ and $D_I\ne0$ in superconductors below the critical temperature. Furthermore an ideal conductor is characterized by a vanishing $D_{I,M}$ but $D_I\ne0$ at finite temperature. 

It should be mentioned that there is an important relationship between the response function $\Psi_I(z)$ and the relaxation function $\Sigma_I(z)$ introduced by Mori (see Appendix \ref{sec:app_transport}). It describes the response of the system, in thermal equilibrium at $t=-\infty$, to a small classical field $h$ (which couples to the observable $I$) applied adiabatically from $t=\
-\infty$ and cut off at $t=0$.

Furthermore, we point out that the dynamical spectra $\Psi_{I,\text{reg}}(\omega)$, the experimentally measurable quantity, can be calculated once the function $\langle I(\tau)I(0)\rangle$ is known ($\tau\ge0$). Indeed, it is possible to show (see Appendix \ref{sec:app_transport}) that there is an exact relation between the function $\Psi_{I,\text{reg}}(\omega)$ and  $\langle I(\tau)I(0)\rangle$: 
\begin{equation}\label{eqn:regular}
    \int_{-\infty}^{+\infty}d\omega \,\frac{\hbar\omega}{2\pi}\Psi_{I,\text{reg}}(\omega)D_{\omega}(\tau) =\langle I(\tau)I(0)\rangle- \frac{D_{I} -D_{I,M}}{\beta}.
\end{equation}
Then, by performing the analytic continuation, for example by using the maximum entropy method\cite{jarrell}, one can extract real frequency, dynamical information from imaginary-time correlation functions computed in quantum Monte Carlo simulations. In fact, we emphasize that, when $\tau\ge0$, $\langle I(\tau)I(0)\rangle$ coincides with $-\hbar \Pi_I(\tau)$, and, furthermore, the function $\Pi_I(\tau)$ allows one to get also $D_I$ and $D_{I,M}$ by using Eq.\ref{eqn:Drude} and Eq.\ref{eqn:Drude1}, respectively\cite{Evertz_PRB}. 

Once the physical quantities $\Psi_{I,\text{reg}}(\omega)$, $D_I$  and $D_{I,M}$ are known, one can determine quantum fluctuations of the $Q$ operator at real times: 
\begin{equation}\label{eqn:dynamic_general_q}
    \begin{split}
        &\frac{d}{dt}\biggl\langle\bigl(Q(t)-Q(0)\bigr)^2\biggr \rangle = \frac{2(D_I-D_{I,M})t}{\beta} \\&+\int_0^{+\infty} d\omega \, \frac{2\hbar\Psi_{I,\text{reg}}(\omega) \sin(\omega t)}{\pi\tanh(\beta\hbar\omega/2)},
    \end{split}
\end{equation}
i.e. the instantaneous charge diffusivity.   

Now we introduce the second biasing scheme that involves another response function: $\Psi_Q(z)$, i.e the observable $A$ is the charge operator $Q$. Let us start by adding a small current bias term $-I(t)\phi$ into the Hamiltonian, so that $H_{tot}=H-I(t)\phi$, where $I(t)$  is an assigned function of the time ($t>0$, as above for $t<0$ the system is in thermodynamic equilibrium).  The next step is to perform a gauge transformation. If $\chi(t)$ is the solution of $i\hbar\frac{\partial\chi}{\partial t}=H_{tot}\chi(t)$, letting $\chi(t)=e^{-S}\chi^{\prime}$, where $S=-i \frac{\phi}{\hbar}f(t)$ and $f(t)=\int I(t) dt$, $\chi^{\prime}(t)$ is the solution of the following Schr\"odinger equation: $i\hbar\frac{\partial\chi^{\prime}}{\partial t}=H^{\prime}_{tot}\chi(t)$, where $H^{\prime}_{tot}$ is obtained by replacing the charging term $\frac{Q^2}{2C}$ in $H$ with $\frac{(Q+f(t))^2}{2C}$. In other words, after the gauge transformation, the contribution $-I(t)\phi$ is missing and the charge operator $Q$ is now represented by the operator $\tilde{Q} =Q+f(t)$. Now, following standard textbook procedure\cite{Mahan, Fetter}, we apply the linear response theory to the transformed charge operator getting for the Fourier transforms:
\begin{equation}\label{eqn:qtz}
\tilde{Q}(z)=\frac{I(z)}{C}\frac{i}{z}\bigl(C+\Pi_Q(z)\bigr),
\end{equation}
i.e $\tilde{Q}(z)=\frac{I(z)}{C}\Psi_{Q}(z)$. This important relation shows that $\Psi_{Q}(z)$ allows us to determine the voltage across the junction $\frac{\tilde{Q}(z)}{C}$ when a small current bias term is included in the Hamiltonian.  Then it provides another way of characterizing the transport properties of the junction. In the mechanic analog, $I(z)$ plays the role of the electric field, $\frac{\tilde{Q}}{C}$ represents the total current, and $\frac{\Pi_Q(z)}{C^2}$ is the current-current correlation function. The corresponding response function provides the well-known Kubo formula for the conductivity of electrons in the continuum approximation: $\sigma(z)=\frac{i}{z}\bigl(\frac{n e^2}{m}+\Pi(z)\bigr),$ $n$ being the density of the charge carriers. Then the quantities $D_Q$ and $D_{Q,M}$ allow us to discriminate between insulators, superconductors, ideal conductors, and conductors. For this reason, in many papers on this topic, the main objective has been the calculation of the mobility of the phase particle. On the other hand, it is worth emphasizing that in this second biasing scheme, the presence of a delta function centered at $\omega=0$, i.e. a nonvanishing value of $D_Q$, indicates that the system is an insulator. In fact, the mobility of the phase particle corresponds to the resistance of the junction.  

In ref.\cite{microwave} the effective admittance of the junction, in the case of the CL model, has been addressed at small values of $\alpha$ in the transmon regime ($E_J/E_C\gg 1$) and, by invoking the duality relation, in the strong coupling regime of the charge qubit ($E_J/E_C\ll 1$). The calculation has been perturbatively performed by starting from $\alpha=0$ and mapping the Hamiltonian on an effective sine-Gordon model. Here we derive the linear response of the junction, in both models known in the literature, by using a numerical approach based on the maximum entropy method combined with the Monte Carlo technique on the imaginary axis. It allows us to perform the analytic continuation on the real axis. Our attention will be focused, in particular, on the intermediate regime: $E_J/E_C=0.5$. 

Once the physical quantities $\Psi_{Q,\text{reg}}(\omega)$, $D_Q$  and $D_{Q,M}$ are known, one can determine quantum fluctuations of the $\Phi$ operator at real times: 
\begin{equation}\label{eqn:dynamic_general1}
    \begin{split}
        &\frac{d}{dt}\biggl\langle\bigl(\phi(t)-\phi(0)\bigr)^2\biggr \rangle =\frac{2(D_Q-D_{Q,M})t}{\beta C^2} \\&+\int_0^{+\infty} d\omega \, \frac{2\hbar\Psi_{Q,\text{reg}}(\omega) \sin(\omega t)}{\pi C^2\tanh(\beta\hbar\omega/2)},
    \end{split}
\end{equation}
i.e. the instantaneous flux diffusivity.   

In the following, we will show that, within the linear response theory, physics dramatically changes depending on whether the phase fluctuations are included or not in the Hamiltonian. 

\section{Classical dissipative junction}
We first analyze the transport properties of the classical dissipative Josephson junction, i.e. $C \to \infty$, within both the QP and CL models. In other terms, the phase is locked as in the pioneering works on this subject; on the other hand, we retain the effects introduced by subgap and shunt resistances. This simplified model is exactly solvable. 

Let us start our discussion with the QP Hamiltonian.  It can be easily diagonalized by performing a unitary transformation,   $\bar{H}=e^{\bar{S}}H_{QP}e^{-\bar{S}}$, where $\bar{S} = \sum_{i=1}^N\sum_{j=1}^2\frac{S_{i,j}}{\hbar\omega_{i,j}}\sqrt{\frac{\hbar}{2C_{i,j}\omega_{i,j}}}(a^{\dagger}_{i,j}-a_{i,j})f_j(\frac{\phi}{\phi_0}\frac{2\pi}{2})$, and $a_{i,j}$ and $a^{\dagger}_{i,j}$ are the annihilation and creation operators associated with the two bosonic baths ($\phi_{i,j}=\sqrt{\frac{\hbar}{2C_{i,j}\omega_{i,j}}}(a^{\dagger}_{i,j}+a_{i,j})$) and $f_1(x)=\cos(x)$, $f_2(x)=\sin(x)$. The transformed Hamiltonian,  $\bar{H}=e^{\bar{S}}H_{QP}e^{-\bar{S}}$, now consists of three non-interacting fields:   
\begin{equation}\label{eqn:diagonalized}
    \bar{H}=-E_J \cos\biggl(\frac{\phi}{\phi_0}2\pi\biggr)+\sum_{i=1}^N \sum_{j=1}^2\hbar\omega_{i,j}a^{\dagger}_{i,j}a_{i,j}+c,
\end{equation}
where the constant $c$ is given by $c=-\frac{2\hbar^2}{\pi^2}\int d\omega \frac{J(\omega)}{\hbar \omega}$. In particular, the phase $\phi$ is a constant of motion. The ground state is degenerate and corresponds to the phase particle located in one of the minima of the potential $-E_J \cos\biggl(\frac{\phi}{\phi_0}2\pi\biggr)$.  By performing the same unitary transformation on the current operator, it is straightforward to prove that: $D_I = E_J (2e/\hbar)^2\langle\cos(2\pi\phi/\phi_0)\rangle$ and $D_{I,M} = E_J (2e/\hbar)^2\bigl[\langle\cos(2\pi\phi/\phi_0)\rangle-\beta E_J\langle\sin^2(2\pi\phi/\phi_0)\rangle\bigl]$, where $\langle A \rangle = \int d \phi \, e^{\beta E_J\cos(2\pi\phi/\phi_0)}A(\phi)/\int d\phi \, e^{\beta E_J\cos(2\pi\phi/\phi_0)}$.  The integrations can be exactly performed: the calculation yields $D_{I,M} = 0$ and $D_I= E_J (2e/\hbar)^2 \frac{I_1(\beta E_J)}{I_0(\beta E_J)}$, where $I_n(x)$ is the modified Bessel of the first kind: $D_I$ turns out to be independent of $ \alpha$ and decreases with increasing temperature. Finally, $Y_{\text{reg}}(\omega) = 1/R_{QP} $ for $-\omega_D<\omega<\omega_D$, i.e. $Y_{\text{reg}}(\omega)$ is constant as a function of $\omega$. 

These results are exact and deserve particular attention. Being $D_{I,M}=0$ and $D_I\ne0$, one is tempted to conclude that the classical Josephson junction is an ideal conductor, that is the result found by Josephson. On the other hand, the regular part of the conductivity does not exhibit any gap. In general, in a BCS superconductor, the optical absorption shows a delta function centered at $\omega=0$ and a gap $2 \Delta$, $\Delta$ being the energy gap in the density of states. The absence of a gap in the excitation spectra of the current-current correlation function has significant consequences for the motion of the charges. Indeed, by using Eq.\ref{eqn:dynamic_general_q}, it is straightforward to show that at long times, $t\gg \beta\hbar$, the charge fluctuations in thermal equilibrium, in addition to a dissipationless motion stemming from a non-vanishing $D_I$, exhibit also a diffusive contribution: $\frac{\langle (Q(t)-Q(0))^2\rangle}{e^2}=\frac{D_I t^2}{e^2 \beta}+\frac{4 \alpha}{\pi}\frac{t}{\beta\hbar}$. It implies that, in the presence of non-vanishing subgap conductance, $G_{QP}=1/R_{QP}$, no matter how small, the classical Josephson junction beside ideal conduction (being $D_I \ne 0$) exhibits also a typical dissipative term (diffusive motion of the charges), i.e. the ideal conduction is not protected due to a finite value of the regular part when $\omega \rightarrow 0$. It resembles the behavior of an ideal bose gas where the critical velocity is zero: as soon as particles flow along a capillary at finite velocity, no matter how small, viscosity sets in. 

In the following, we will show that the quantum phase fluctuations completely modify this scenario: a gap opens in the spectra and, at the same time, the weight of the Drude term vanishes, making the junction insulator in the linear response regime. In other words, $E_C=0$ is different from $E_C\rightarrow 0$, i.e. a non-analytic behavior occurs as a function of $E_C$. Then the main message is that the ideal conduction, foreseen in the pioneering papers on this topic, is not robust against neither quantum phase fluctuations nor the dissipative effects. 

We emphasize that the above achieved conclusions are independent of the details of the model. In fact, the same results can be obtained within the CL model. By defining $\bar{S} = \sum_{i=1}^N \frac{g_i}{\hbar\omega_i}(a_i-a^{\dagger}_i)\phi$, where $g_i = \frac{1}{L_i}\sqrt{\frac{\hbar}{2C_i\omega_i}}$ and $a_i$ $(a^\dagger_i)$ represents the annihilation (creation) operator for the bosonic bath, the unitary transformation, $\bar{H}=e^{\bar{S}}H_{CL}e^{-\bar{S}}$, leads to $D_{I,C} =0$, $D_I = E_J (2e/\hbar)^2\langle\cos(2\pi\phi/\phi_0)\rangle$, and $Y_{\text{reg}}(\omega) = 1/R_{S} $ for $-\omega_D<\omega<\omega_D$. Then, we emphasize that both the QP and CL models for the classical Josephson junction yield identical results at the level of linear transport. Notably, in the QP model, the subgap resistance plays the same role as the shunt resistance in the CL model. 

\section{The Cooper pair box}
Here we clarify the effects induced by the quantum phase fluctuations when the subgap resistance does not play any role in the parallel ($R_{QP} \rightarrow \infty$, i.e. $G_{QP}=0$) and the junction is not shunted, i.e. the so-called Cooper pair box. In this specific case, the Hamiltonian turns out to be: $H_{JJ}=\frac{Q^2}{2C}-E_J \cos[\frac{\phi}{ \phi_0}2\pi]$. 

The calculations can be performed on a very large system with periodic boundary conditions by using an exact diagonalization approach. In the mechanic analog, this Hamiltonian describes the motion of a fictitious particle in a periodic potential where the lattice parameter is $\phi_0$. We use the basis states: $e^{i Q \phi}/\sqrt{N\phi_0}$. Here N (even) represents the number of cells adopted, each of them of length $\phi_0$ and $Q=\frac{2\pi}{\phi_0}(\frac{n}{N}+r)$, where $n$ and $r$ are two integers, $n$ ranging from $-N/2+1$ to $N/2$. These wavefunctions are the eigenstates of the Hamiltonian in the absence of the periodic potential. On the other hand, when $E_J\ne0$, these basis states are coupled, but it is straightforward to prove that $n$ is again a good quantum number. In this case, the eigenstates are the Bloch wavefunctions: $\Psi_{k,s}(\phi)$, where $k=\frac{2\pi}{\phi_0}\frac{n}{N}$ lies within the first Brillouin zone and $s$ is the band index. One obtains the well-known band structure of the solid state theory. The number of cells $N$ and the number of bands used in numerical calculations are increased to obtain the convergence for the specific physical quantity investigated.

Once the exact eigenstates of the Hamiltonian are known, one can determine $\Psi_{A,\text{reg}}(\omega)$, $D_A$, and $D_{A,M}$ for $A=I$ and $A=Q$. In particular, we get $D_{I,M}=D_{I}=D_{Q,M}=0$, while $D_Q\ne0$. These results clearly show that the ideal unshunted quantum Josephson junction is an insulator in the linear response theory, a result that is very different from that obtained in the absence of phase quantum fluctuations.

\begin{figure}
\flushleft
        \includegraphics[scale=0.34]{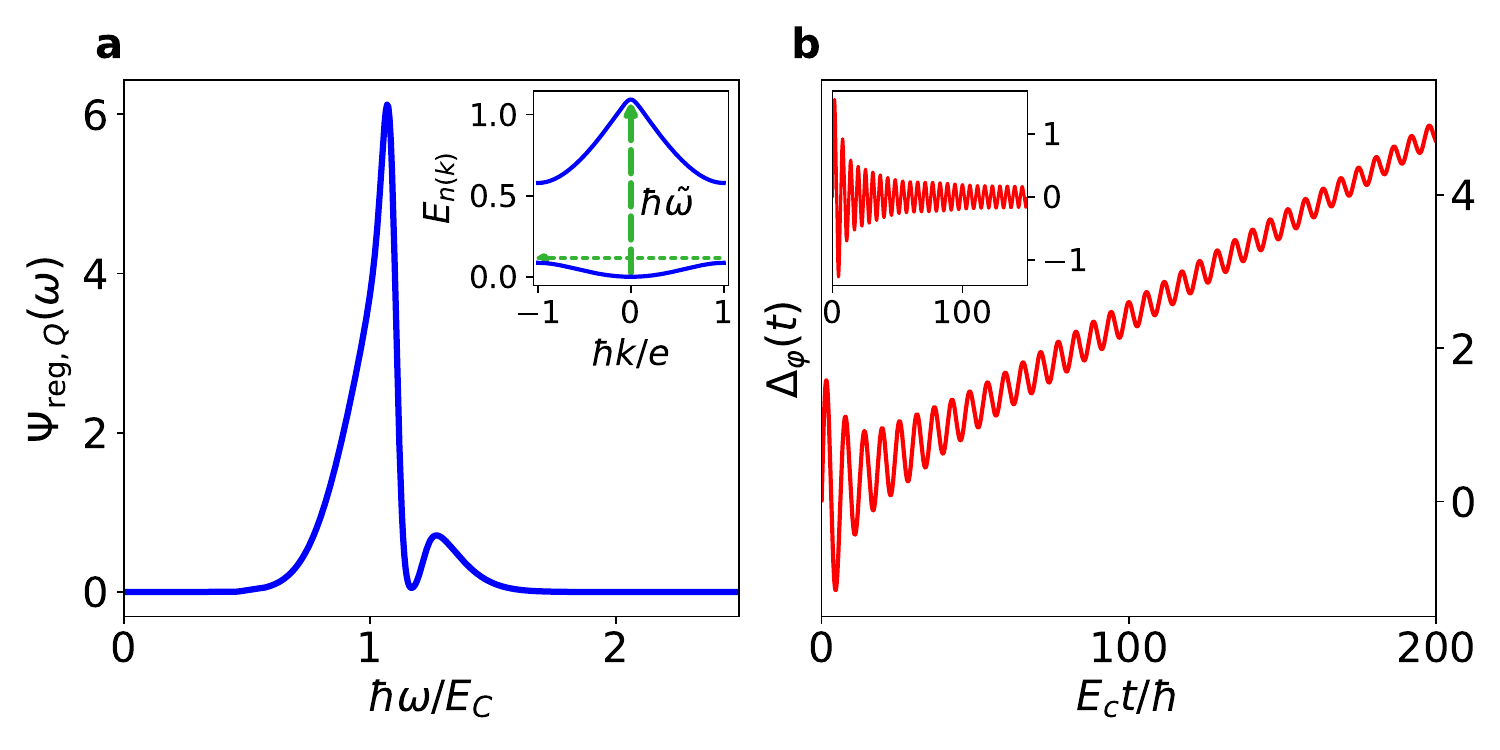}
        \caption{The Cooper pair box: (a) the regular part of the charge response function (in units of $e^2\hbar/E_c^2$ ) as function of $\frac{\hbar \omega}{E_c}$ at $\beta E_c=100$ and $E_J/E_C=0.5$; in the inset the first two energy bands (in units of $E_C$): vertical dashed green line indicates the allowed transition at $\alpha=0$ (corresponding to the frequency $\tilde{\omega}$), horizontal green dotted line represents the umklapp process in the Bloch oscillations; (b) instantaneous phase diffusivity (in units of $\hbar/E_C$) as function of the time; in the inset the sub-diffusive term of $\Delta_{\varphi}(t)$.} 
\label{fig:1}
\end{figure}

In Fig.\ref{fig:1}a we plot the regular part of the charge relaxation function: $\Psi_{Q,\text{reg}}(\omega)$. The main peak is centered at the frequency $\tilde{\omega}$ that, at $T=0$, goes from $E_C$, for $E_J \rightarrow 0$, to the plasma frequency $\omega_{pl}$, when $E_J \rightarrow \infty$. Here, $\omega_{pl}=\sqrt{2 E_J E_C}$ corresponds to the oscillation frequency of the phase particle at the bottom of one of the minima of the potential. This main peak stems from a vertical transition, i.e. the value of $k$ does not change during the transition between the first two lowest energy bands. Indeed, the charge operator does not allow intraband transitions (see inset Fig.\ref{fig:1}a). 
Since it is possible to prove that $Y_{\text{reg}}(\omega)=\omega^2\Psi_{Q,\text{reg}}(\omega)$, the same conclusions hold for the regular part of the admittance $Y_{\text{reg}}(\omega)$. 

We emphasize the presence of a gap in both spectra $Y_{\text{reg}}(\omega)$ and $\Psi_{Q,\text{reg}}(\omega)$, so that the phase and charge mobilities, i.e.  $Y_{\text{reg}}(\omega\rightarrow 0)$ and $\Psi_{Q,\text{reg}}(\omega\rightarrow 0)$, are both vanishing. This implies the absence of diffusive behavior in both phases and charge fluctuations. In particular, phase fluctuations in the thermodynamic equilibrium, $\frac{\langle (\phi(t)-\phi(0))^2\rangle}{(\phi_0/2\pi)^2}$, turn out to be the sum of two different terms:  the first one is ballistic, being $D_Q\ne0$, whereas the other one, stemming from $\Psi_{Q,\text{reg}}(\omega)$, is subdiffusive: in Fig. \ref{fig:1}b we plot the time derivative, i.e. the instantaneous phase diffusivity: $\Delta_{\varphi}(t)=\frac{d \langle(\varphi(t)-\varphi(0))^2\rangle}{dt}$, that turns out to be the sum of a term that increases linearly with time and an oscillating contribution (see inset) with period $\frac{2\pi}{\tilde{\omega}}$ and amplitude decreasing over time depending on the temperature: the higher temperature, the more rapidly the amplitude decreases over time. On the other hand, charge fluctuations, $\langle (Q(t)-Q(0))^2\rangle$, display only subdiffusive behavior, being $D_I=0$. We emphasize that all these results are either analytically or numerically exact.  

Let us briefly pause to summarize what we have done. We found that, in the ideal unshunted case ($G_{QP}=0$), the classical junction is an ideal conductor, whereas the presence of phase quantum fluctuations prevents ideal conduction in the current channel ($D_I=0$ and subdiffusive current behavior) and allows dissipationless behavior in the phase channel: presence of delta function centered at $\omega=0$ with strength $D_Q\ne0$ and gap in the spectra $\Psi_{Q,\text{reg}}(\omega)$. Is then not possible to observe the ideal conduction of the junction, effect predicted by Josephson in 1962? 

The answer has to be sought by investigating beyond the linear response regime: the Bloch oscillations. Indeed, in the presence of current bias $I$ (in the Hamiltonian a term $-I\phi$ appears), a very good approximation can be obtained by solving the semiclassical motion equation for the variable $k$: $\hbar\dot{k}=I$. It is well known, from the solid state theory, that this equation is valid under the assumption that the current bias is too weak to induce transitions from one band to another one, i.e. Zener tunneling is neglected (in the mechanic analog, current plays the role of electric field). In this case, the wave packet accelerates until it reaches the Brillouin zone boundary and jumps to the opposite border of the Brillouin zone. Every time such umklapp event takes place, a Cooper pair tunnels through the junction. Indeed, from $k=\frac{2\pi}{\phi_0}\frac{n}{N}$, it follows $\hbar k=2e  \frac{n}{N}$ and then, at the edges of the Brillouin zone ($n=N/2$ and $n=-N/2+1$), the quasimomentum assumes the values $+e$  and $-e$ ($N\rightarrow\infty$). At the same time, the group velocity of the wavepacket is given by $\frac{1}{\hbar} \frac{\partial E_{k,0}}{\partial k}$, i.e. the voltage across the junction oscillates around zero, whereas the average value of the Cooper pair current is different from zero. In other terms, the junction shows ideal conductivity as a response to a weak dc current bias in the non-linear response regime:  an applied dc current causes voltage oscillations across the junction, with frequency $\frac{I}{2e}$. 

Then, the next and more relevant question is: how is this physics modified when either QP resistance $R_{QP}$ or shunt resistance $R_S$ are taken into account?  To this aim, we will study the full Hamiltonians $H_{QP}$ and $H_{CL}$. First, we will study the thermodynamic equilibrium by using a Monte Carlo technique: we will prove the existence of quantum phase transitions induced by varying the strength of the coupling with the bosonic baths. Second, we will investigate the linear response regime and the related consequences on the semiclassical motion equations beyond the linear response regime.     
\section{The dissipative quantum Josephson junction: thermodynamical properties}

Here we investigate the equilibrium effects induced by both quantum phase fluctuations and dissipation by using a worldline Monte Carlo approach (for details on the method, see Appendix \ref{sec:MQ}). In this context, the resistance of the subgap $R_{QP}$ and the resistance of the shunt $R_S$ play a key role in the QP and CL models, respectively. 

We first focus our attention on the QP model, where dissipation gives rise to a long-range interaction in \eqref{eqn:action_QP}, given by $S_R(\tau,\tau') \propto\sin^2\bigl(\frac{\varphi(\tau)-\varphi(\tau')}{4}\bigr)$. It should be emphasized that this interaction is characterized by a period of $4\pi$. This implies that, if we start from a constant path $\varphi(\tau)$, shifting a segment by an integer multiple of $4\pi$ leaves the dissipative contribution to the total action unchanged, whereas shifting it by a half period results in maximizing its contribution. As a consequence, when $\alpha$ is sufficiently large, at zero temperature, we expect the phase particle to be localized in the even or odd minima of the Josephson potential. On the other hand, in the absence of dissipation, i.e. $\alpha = 0$, the ground state predicts a delocalized state. This suggests that the system undergoes a quantum phase transition (QPT) as $\alpha$ is tuned. Moreover, because of the ohmic nature of the dissipation, i.e. $J(\omega) \propto \omega$ up to the cutoff frequency, the kernel of the retarded interaction $K(\tau)$ decays as an inverse square power law in $\tau$ at long imaginary times. Then we foresee the QPT transition to be in the BKT universality class \cite{Grazia_NatComm, Giulio_PRL, Giulio_manyspin_PRB,Giulio_spinboson_PRB}. 

In order to accurately characterize QPT, it is useful to introduce the order parameter $m^2 = \frac{1}{\beta\hbar}\int_0^{\beta\hbar} d\tau \,\langle \cos(\varphi(\tau)/2)\cos(\varphi(0)/2)\rangle$. This parameter vanishes in the delocalized phase, whereas it becomes non-zero as soon as the symmetry between even and odd minima is broken. Moreover, it continuously increases and saturates to 1 when the system is fully localized in even or odd minima.  

\begin{figure}
\centering
        \includegraphics[scale=0.22]{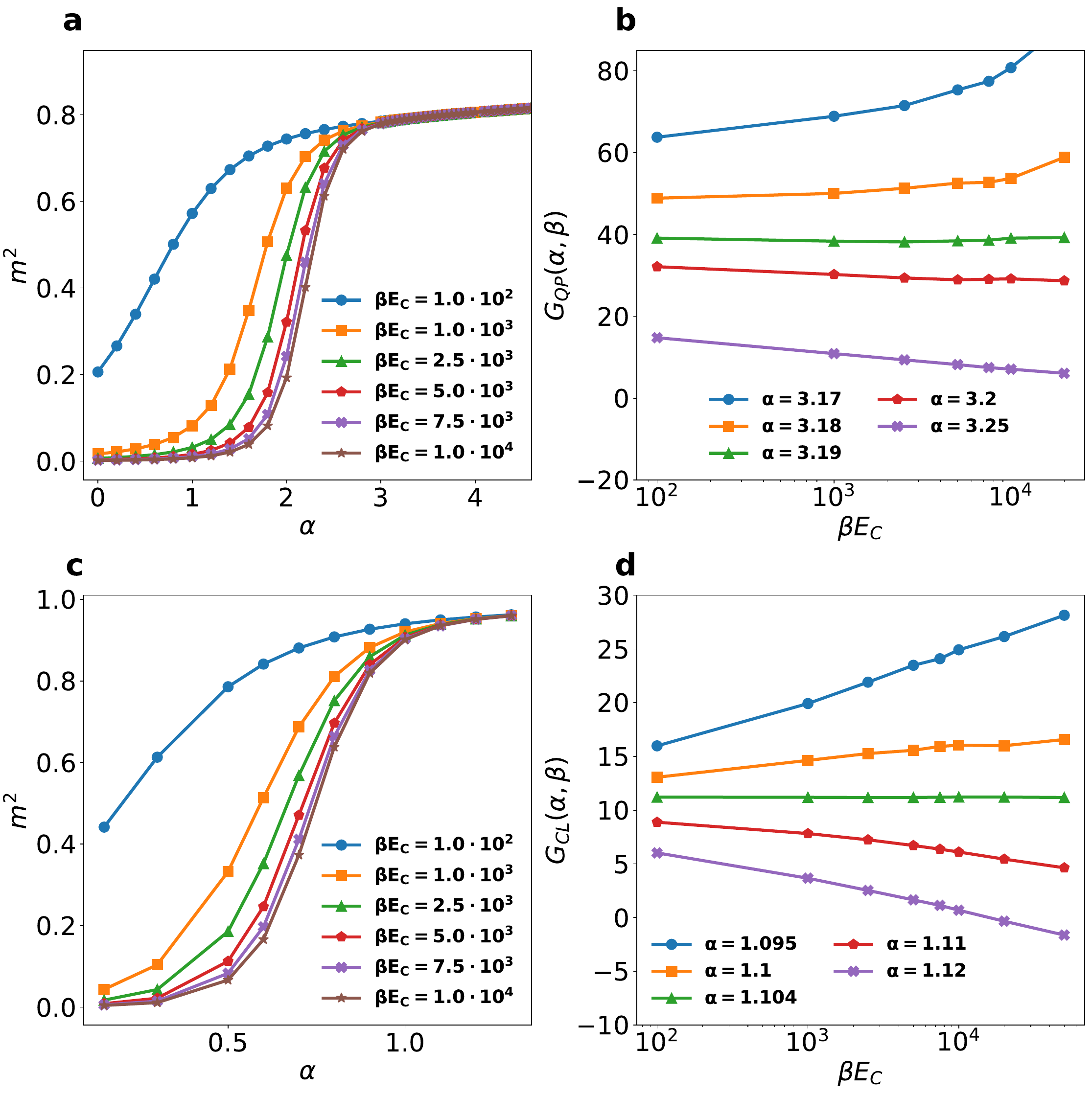}
        \caption{Dissipative Josephson junction: (a) order parameter $m^2$ as a function of $\alpha$ for the QP model ($E_J/E_C=0.5$), with $\beta$ ranging from $\beta_{\min}E_c=100$ to $\beta_{\max}E_c = 10000$; (b) scaling function $G_{QP}(\alpha, \beta)$ as a function of $\beta$ for values of $\alpha$ near the candidate critical point $\alpha_c$; (c) order parameter $m^2$ as a function of $\alpha$ for the CL model, with $\beta$ ranging from $\beta_{\min}E_c=100$ to $\beta_{\max}E_c = 10000$; (d) scaling function $G_{CL}(\alpha, \beta)$ as a function of $\beta$ for values of $\alpha$ near the candidate critical point $\alpha_c$. We set $\Psi_{CL,c} = 1$ in our analysis.} 
\label{fig:2}
\end{figure}

In Fig.\ref{fig:2}a we plot the order parameter $m^2$ at $E_J/E_C = 0.5$ as a function of $\alpha$ for different temperatures ranging from $k_B T = 10^{-2} E_C$ to $k_B T = 10^{-4} E_C$. As expected, the curves increase and become progressively steeper as $\beta = \frac{1}{K_B T}$ grows. This behavior is one of the signatures of the BKT transition, since BKT theory predicts that $m^2$ exhibits a jump discontinuity at a critical value $\alpha_c$ at zero temperature. To clarify the universality class of the QPT, we analyze the scaling properties of the order parameter. Defining $\alpha_{\text{eff}} = \frac{4\hbar\alpha}{\pi^2}$, the action takes the form:

\begin{equation*}
    \begin{split}
        &\mathcal{S}[\varphi(\tau)] = \mathcal{S}_{JJ}[\varphi(\tau)] \\ +&\frac{1}{2}\int_0^{\beta\hbar} d\tau \, \int_0^{\beta\hbar} d\tau' \, \cos\bigl(\varphi(\tau)/2\bigr) K_{\text{eff}}(\tau -\tau') \cos\bigl(\varphi(\tau')/2\bigr)  \\
        +&\frac{1}{2}\int_0^{\beta\hbar} d\tau \, \int_0^{\beta\hbar} d\tau' \, \sin\bigl(\varphi(\tau)/2\bigr) K_{\text{eff}}(\tau -\tau') \sin\bigl(\varphi(\tau')/2\bigr),
    \end{split}
\end{equation*}

 where the effective kernel has an asymptotic behavior $K_{\text{eff}}(\tau) = \frac{\alpha_{\text{eff}}}{2\tau^2}$ and $S_{JJ}[\varphi(\tau)]$ is the action of the ideal quantum junction. It is worth highlighting that the second term in $\mathcal{S}[\varphi(\tau)]$ is solely responsible for the QPT, since, for $E_J/E_C \neq 0$, the system is expected to be localized around the minima of the Josephson potential rather than the maxima. 

 The order parameter and the term driving the QPT closely resemble those governing the BKT transition in the spin-boson model, where the spin variable is replaced by $\cos(\varphi/2)$, which assumes values $\pm1$ in the minima of the potential. For this reason, we adopt the same scaling argument proposed for spin-boson~\cite{Giulio_PRL, Giulio_manyspin_PRB, Giulio_spinboson_PRB}. We introduce the scaling function $\Psi_{QP}(\alpha,\beta ) = \alpha_{\text{eff}}m^2$, which, for large $\beta$, follows the asymptotical behavior $\frac{\Psi_{QP}(\alpha_c, \beta)}{\Psi_{QP, c}} = 1+ \frac{1}{2(\ln\beta-\ln \beta_0)}$ in the BKT framework, where $\Psi_{QP,c} = \Psi_{QP}(\alpha_c, \beta \to \infty) = 1$ is the universal jump of the scaling function and $\beta_0$ a fitting parameter~\cite{Minn_PRB1, Minn_PRB2, Minn_PRL}. The value of the jump is expected because of the analogy with the spin-boson model, where the same universal jump occurs at criticality.  As a consequence, the function $G_{QP}(\alpha, \beta) = \frac{1}{\Psi_{QP}(\alpha, \beta)-1}-2\ln\beta$ should be independent of $\beta$ at very low temperatures at criticality. By employing this criterion, we estimate $\alpha_c \approx 3.19$, as $G_{QP}(\alpha, \beta)$ appears to become independent of $\beta$ (see Fig.\ref{fig:2}b). 
 
 We emphasize that, above $\alpha_c$, there is no complete localization: the phase particle is smeared either for all even or odd minima of the periodic potential: only the tunneling to the nearest minimum is completely suppressed. On the other hand, the amplitude of tunneling between two nearest even or odd minima is finite. Korshunov\cite{korshunov} has explicitly calculated the bandwidth at large couplings and $T=0$, finding that it goes exponentially to zero as a function of the strength of the interaction with the baths. At $E_J/E_C=0.5$ and $\alpha_c=3.19$, the bandwidth turns out to be of the order of $10^{-8} E_C$. Due to this very small value of the effective bandwidth, at $T \ne 0$, physical effects described by the QP Hamiltonian will be very similar to those corresponding to a complete localization, as will be shown below.    

 On the other hand, within the CL model the localization is expected to be complete even at $T=0$\cite{Schmid, Bulgadaev, Troyer_MC}: the phase particle, above a critical value of the coupling with the bath $\alpha_c$, is localized in one of the minima of potential $-E_J \cos(\varphi)$. The critical value of $\alpha$, $\alpha_c$, is expected to be $1$, independent of the value of the ratio $\frac{E_J}{E_C}$, in the limit $\omega_D \rightarrow \infty$. Our calculations, performed at $\frac{E_J}{E_C}=0.5$ and $\omega_D=50 E_C$, agree with this prediction and allow us to characterize the nature of this QPT.   
 
 First, we partition the real axis into intervals $[(2n-1)\pi,(2n+1)\pi)$, with $n\in{\mathbb{Z}}$. Each of these intervals represents a basin of attraction of one of the minima of the cosine potential. Given a value $\phi(\tau)$ of the phase in imaginary time $\tau$, the index $n$ of the basin to which it belongs is given by $\nint\left[\frac{\phi(\tau)}{2\pi}\right]$, where $\nint\left[x\right]$ is the integer nearest to $x$. Then we define the variable 
 \begin{equation*}
     \sigma(\tau)=(-1)^{\nint\left[\frac{\phi(\tau)}{2\pi}\right]},
 \end{equation*} 
 i.e. we map the worldline $\phi(\tau)$ to a step function that is $\sigma(\tau)=1$ if $\phi(\tau)$ belongs to an even minimum, $\sigma(\tau)=-1$ if it belongs to an odd minimum of the potential. The order parameter is given by
 \begin{equation*}
m^2=\frac{1}{\beta\hbar}\int\limits_0^{ \beta\hbar}\langle\sigma(\tau)\sigma(0)\rangle\,d\tau.
\end{equation*}

In this way, we have mapped the worldline on a sequence of instantons and anti-instantons. In other terms, the function $\sigma(\tau)$ allows us to highlight localization effects, disregarding the effects of phase fluctuations around the minima of potential. 

 As illustrated in Fig.\ref{fig:2}c, the order parameter behaves similarly to that of the QP model. To characterize the universality class of this phase transition, we introduce the scaling function $\Psi_{CL}(\alpha, \beta) = \alpha m^2$. As in the QP case, this function follows the same asymptotic behavior within the BKT framework. In fact, the function $G_{CL}(\alpha, \beta) = \frac{1}{\Psi_{CL}(\alpha,\beta)/\Psi_{CL,c} -1} -2\ln\beta$ should exhibit the same behavior as $G_{QP}(\alpha, \beta)$ at criticality. Following this approach, we find $\alpha_c \approx  1$: $\alpha_c=1.104$ and $\Psi_{CL, c} = 1$ (see Fig.~\ref{fig:2}d), i.e. the jump is universal. The small difference between the estimated critical value of the coupling and the value predicted in the literature is due to the finite value of $\omega_D$ used in our calculations. 

\section{Consequences of QPT on the transport properties}\label{sec:Consequences}
Here, first of all, we recall the basics of the theory predicted by Schmid~\cite{Schmid} and Bulgadaev~\cite{Bulgadaev} in 1983 and 1984, respectively. The bulk of these papers are devoted to the study of the phase particle dc mobility $\mu$, mainly in the CL model. The most important achievement is that the motion is diffusive for $\alpha<\alpha_c$  ($\mu\ne0$) and becomes localized for $\alpha>\alpha_c$  ($\mu=0$). To prove this statement, the behavior of dimensionless mobility $\mu(i\omega_n)=-\frac{\alpha}{2\pi}\omega_n \hbar \Pi_\varphi(i\omega_n)$ has been investigated as a function of $i\omega_n$ for different values of temperature and coupling with bosonic baths. To point out the physical meaning of this procedure,  we notice that, through a double integration by parts, it is straightforward to demonstrate that the following relationship, between phase-phase and charge-charge correlation functions, is fulfilled: $\Pi_\varphi(z)=\frac{1}{z^2}(\frac{1}{C}+\frac{1}{C^2}\Pi_Q(z))$ (see Appendix \ref{sec:Mob}) for more details). By merging it with Eq.\ref{eqn:qtz}, we find that the dimensionless mobility of the phase particle, in the upper half-plane of the complex plane $z$, is given by: $\mu(z)=\frac{1}{R}\frac{V(z)}{I(z)}=\frac{\alpha}{2\pi} i z \hbar \Pi_\varphi(z)$, being $V(z)=\frac{\tilde{Q}(z)}{C}$ ($R$ indicates the shunt (subgap) resistance within the CL (microscopic) model). We recall that in the mechanical analog, $I$, $V$, and $C$, play the roles of the electric field, velocity, and mass, respectively. Then it is clear that, in pioneering works on this topic, the dimensionless mobility of the phase particle has been studied along the imaginary axis ($z=i\omega_n=\frac{2n\pi}{\hbar \beta}$), trying to extrapolate its value at $z\rightarrow0$ in the limit $\beta\rightarrow\infty$. 
The prediction of Schmid, within the CL model, is that $\mu$ goes to $1$ when $\alpha\rightarrow0$ and becomes $0$ for $\alpha>\alpha_c$. Our calculations are completely in agreement with this statement within the CL model. On the other hand, within the microscopic model of the junction, the outcomes of the worldline MC indicate that $\mu$ does not approach $1$ when $\alpha\rightarrow 0$. In any case, what is the main consequence on the transport properties of the finding $\mu\rightarrow 1$ when $\alpha\rightarrow0$? In ref~\cite{Likharev, Schon_Zaikin_review, Herrero_Zaikin} the effect of the Ohmic resistor on the Bloch oscillations, found in the Cooper pair box, has been perturbatively introduced by adding the term $-V/R$ in the motion equation: 

\begin{equation}\label{eqn:motion}
\hbar\dot{k}=I-\frac{V}{R}. 
\end{equation} 

The conclusion is that, being $\mu(z)=\frac{1}{R}\frac{V(z)}{I(z)}$, if dimensionless mobility tends to $1$ when $\alpha\rightarrow0$, at long times ($\omega\rightarrow0$) the right side of Eq.\ref{eqn:motion} goes to zero, so that there is a stationary solution with constant quasicharge and voltage. This implies that the whole bias current passes through the shunt, and the junction is an ideal insulator. The authors go beyond and, by taking into account that the width of the lowest energy bands is finite, suppose that there is a critical value of the bias current, above which the Bloch oscillations start again. 

Since our calculations prove that $\mu$ does not approach $1$ in the weak coupling regime within the microscopic model (see Appendix \ref{sec:Mob}), the first question is: why should this reasoning be true within the CL model and fail in the other Ohmic channel, i.e. in the case $R=R_{QP}$? However, independently of this remark, the main criticism to this theory stems from the observation that the dc mobility, $\mu(\omega\rightarrow0)$, has been evaluated by performing the limit along the imaginary axis. This procedure is well defined when there are no singularities at $z=0$, the origin of the complex plane. It is worth noticing that, in the last four decades, there is no study concerning the presence of non-vanishing Drude weight in the charge response function $\Psi_Q(z)$ (note that the dimensionless mobility of the phase particle in the complex plane is proportional to $\Psi_Q(z)$ (see Appendix \ref{sec:Mob})). So, if $D_Q\ne0$, a term $i\frac{D_Q}{z}$ emerges in the response function $\Psi_{Q}(z)$: the presence of this contribution ensures that the limits $z\rightarrow0$, performed along imaginary and real axes, can lead to two different results. On the other hand, it is well known that the correct physical response is obtained by taking $\Psi_{\text{reg},Q}(\omega\rightarrow 0)$, i.e. from the analytic continuation on the real axis. 
\begin{figure}
\flushleft
        \includegraphics[scale=0.33]{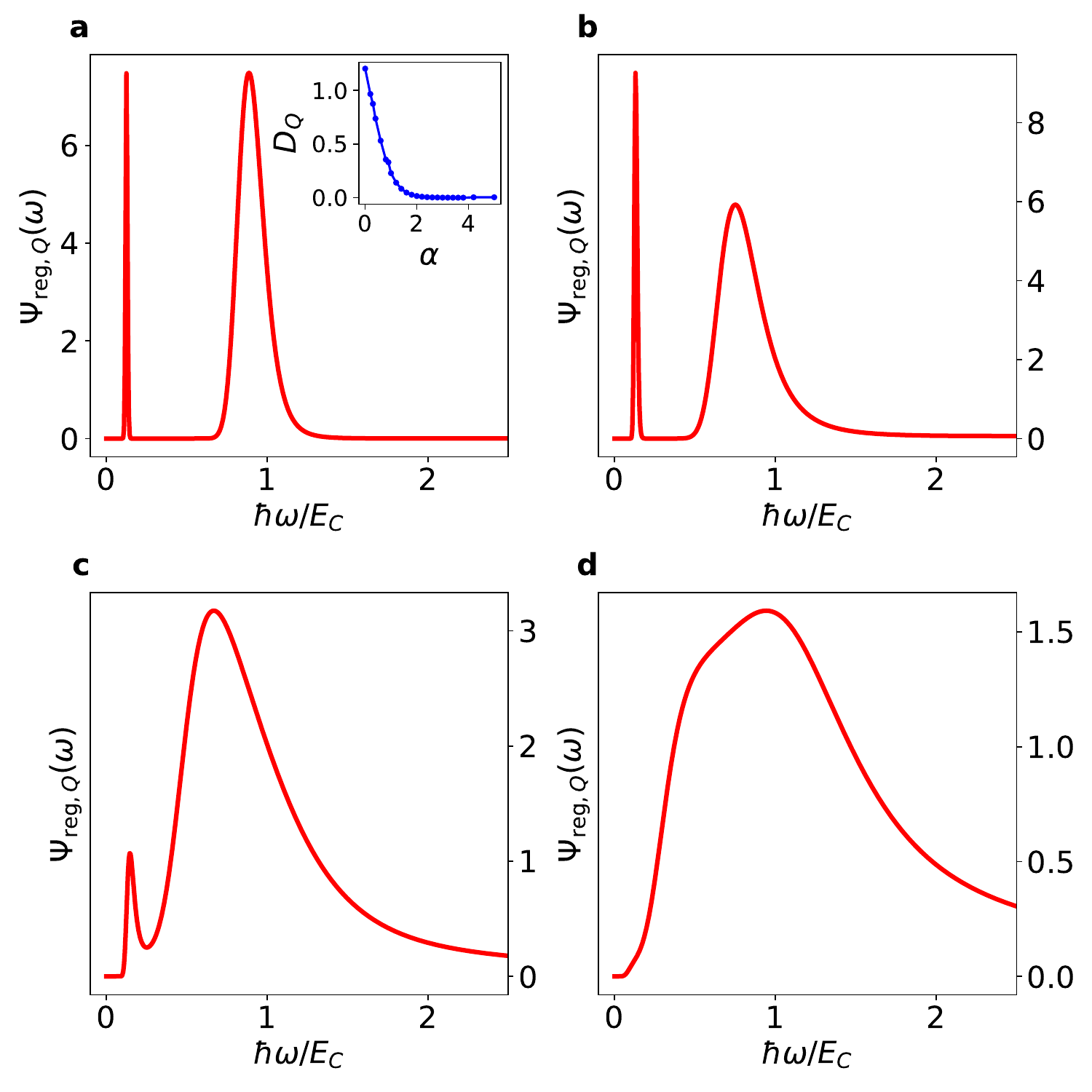}
        \caption{QP model: regular part of the charge response function (in units of $e^2\hbar/E_C^2$ ) as function of $\frac{\hbar \omega}{E_C}$ at $\beta E_C=100$ and $E_J/E_C=0.5$ for $\alpha = 0.3$ (a), $\alpha = 0.9$ (b), $\alpha = 2$ (c), $\alpha = 4.2$ (d); in the inset of (a): charge Drude weight  (in units of $e^2/E_C$) as a function of $\alpha$ at $\beta E_C = 100$.} 
\label{fig:3}
\end{figure}

\begin{figure}
\flushleft
        \includegraphics[scale=0.33]{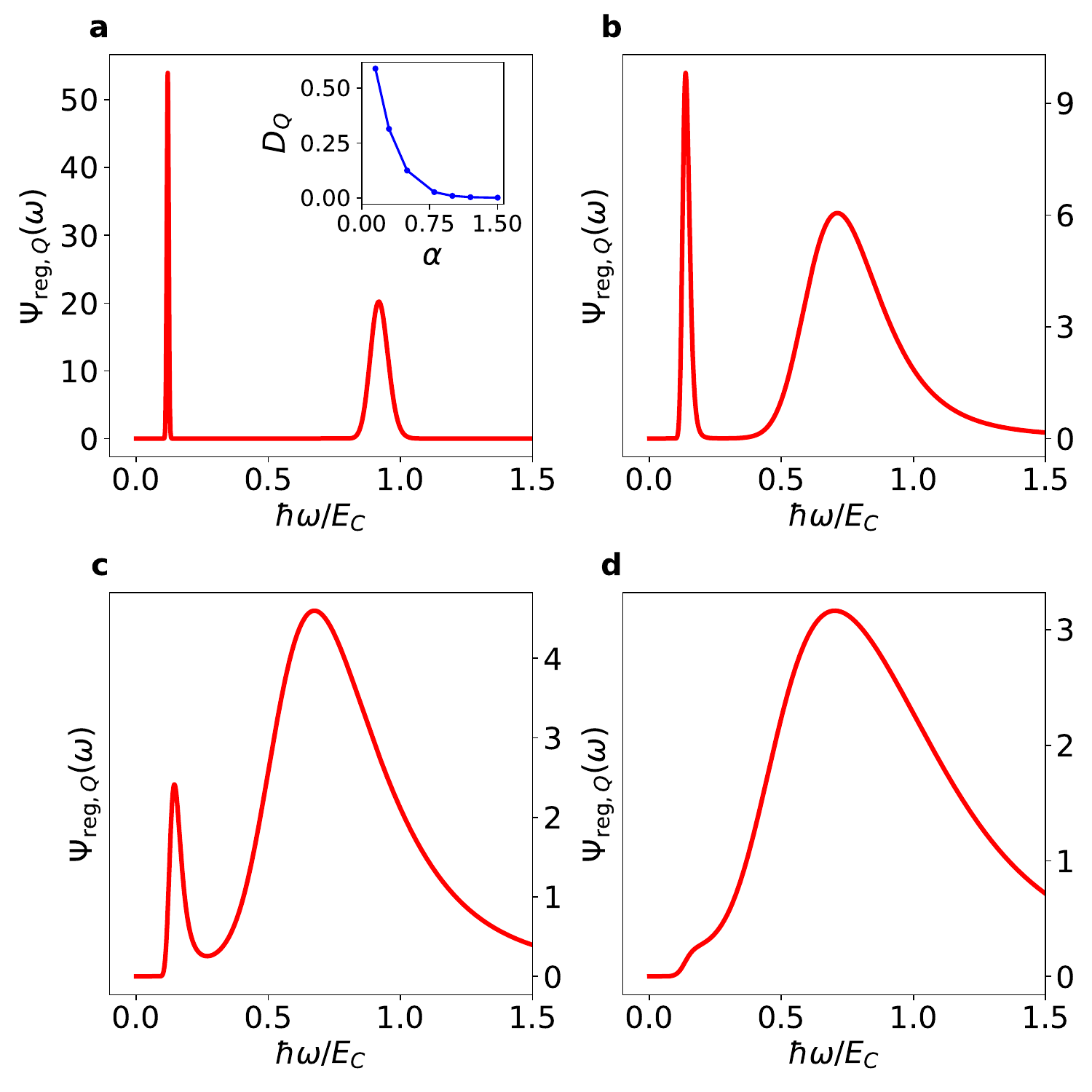}
        \caption{CL model: regular part of the charge response function (in units of $e^2\hbar/E_C^2$ ) as function of $\frac{\hbar \omega}{E_C}$ at $\beta E_C=100$ and $E_J/E_C=0.5$ for $\alpha = 0.15$ (a), $\alpha = 0.5$ (b), $\alpha = 0.8$ (c), $\alpha = 1.5$ (d); in the inset of (a): charge Drude weight (in units of $e^2/E_C$) as a function of $\alpha$ at $\beta E_C = 100$.} 
\label{fig:4}
\end{figure}

\begin{figure}
\flushleft
        \includegraphics[scale=0.34]{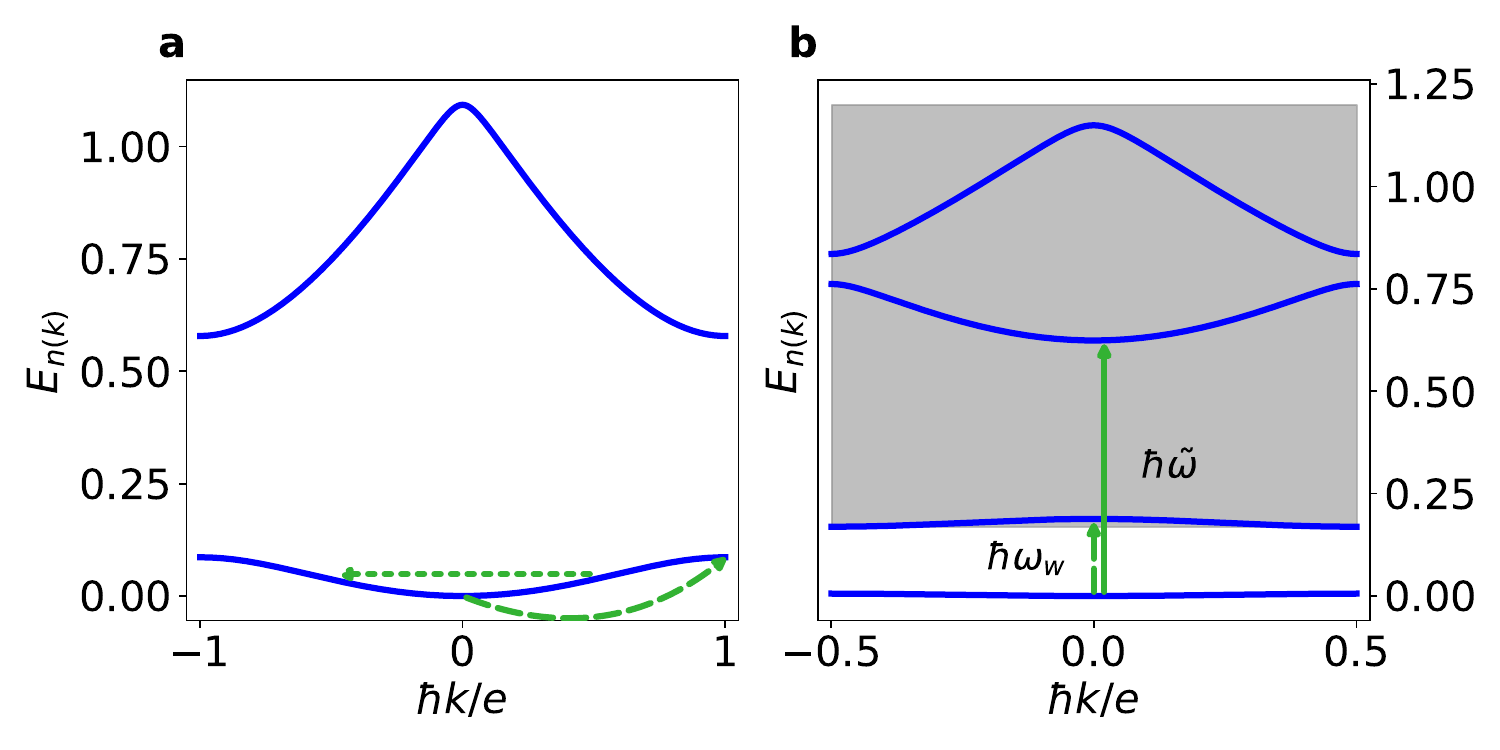}
        \caption{(a) The first two lowest energy bands of the Cooper pair box (in units of $E_C$) at $E_J/E_C=0.5$. The dashed green line indicates the intraband phonon assisted transition ($\alpha\ne0$), whereas the dotted green line is the quasicharge change in a single quasiparticle umklapp event. (b) A sketch of the first 4 energy bands for $\alpha\ne0$ in the reduced Brillouin zone: the vertical lines indicate the two energies corresponding to the two peaks present in $\Psi_{\text{reg},Q}(\omega)$ in Fig.\ref{fig:3} and Fig.\ref{fig:4}. The grey background indicates the presence of a continuum of energies, due to the scattering with phonons of the baths}  
\label{fig:5}
\end{figure}

Our results, achieved by using the maximum entropy method, clearly show that mobility vanishes for any value of coupling $\alpha$ in both models (see Fig.\ref{fig:3} and Fig.\ref{fig:4}). In other words, along the real axis $\mu(\omega\rightarrow0)=0$ for each value of $\alpha$, whereas $\mu(i\omega\rightarrow 0)$ is finite in both the QP and CL models and, in particular, holds $1$ in the weak coupling limit within the $CL$ model. Notably, a delta function, centered at $\omega=0$, is present with weight $D_Q$ (see the insets in panels (a)). It implies that the motion is not diffusive, but ballistic! At any non-vanishing temperature, $D_Q$ decreases with $\alpha$, but is always different from zero within our numerical resolution, although very small even at large values of $\alpha$. In other terms, at $\alpha\ne0$, the features occurring in $\Psi_{\text{reg},Q}(\omega)$ are quite similar to those obtained in the Cooper pair box. The first difference is visible in the weak-coupling regime, where a well-defined excitation, located at a frequency $\omega_w$ of the order of the lowest energy band width within the Cooper pair box, appears in the spectrum. By increasing the strength of the coupling with the bosons, this excitation first becomes a resonance and then is scarcely visible at $\alpha\ge\alpha_c$. It corresponds to a phonon-assisted transition from the minimum ($k=0$) to the maximum ($\hbar k=\pm e$) of the lowest energy band, an intraband transition, non-vertical with respect to the quasicharge $\hbar k$ (see Fig.\ref{fig:5}a). The transferred "momentum" is $\pm e$. It gives rise, at long times, to an oscillating, subdiffusive contribution in the phase fluctuations at the thermodynamic equilibrium. 

Our calculations also show that $D_I=D_{I,M}=0$ and $\Psi_{\text{reg},I}(\omega)$ exhibits a gap, so that, within the linear response regime, independently of the biasing scheme, the junction turns out to be an insulator. Another difference with respect to the typical spectrum of the Cooper pair box is the larger width of the one peak structure in the strong coupling regime: the more $\alpha$ increases, the more the lifetime of the excitation located at $\omega\simeq\tilde{\omega}$ reduces due to scattering with a larger number of bosonic excitations. 

Let us briefly pause to highlight the differences between our findings and scenario present in the literature. In the mechanical analog, one expects that, at $T=0$, a delta function is present in the spectrum of the conductivity (the behavior of the displacement fluctuations is ballistic), whereas, at $T\ne0$, the delta function becomes a Lorentzian, i.e. a Drude peak appears in the spectrum, due to the scattering with phonons,  so that the behavior of the fluctuations switches to diffusive. This is the typical scenario of the polaron physics and it is the underlying principle of the theory proposed by Schmidt in the weak-coupling regime (differently from the polaron, where the motion is always diffusive, here the localization takes place in the strong-coupling regime).

Our calculations prove that the proposed scenario has to be deeply modified. Due to the particular interaction with the environmental degrees of freedom, the behavior of the phase particle fluctuations continues to be ballistic for any coupling, at finite temperature, exactly the same as $\alpha=0$. Furthermore, an energy gap characterizes the spectra: the first excitation is found at a frequency corresponding to the width of the lowest energy band of the Cooper pair box. Since fluctuations at the thermodynamic equilibrium are strictly connected to the response of the system to a weak disturbance, it is natural to wonder which is the effect of this new excitation on the Bloch oscillations observed in the Cooper pair box. 

To this aim, we point out that, within the microscopic model, the interaction with bosonic baths induces intraband scatterings with transferred quasicharge $\hbar k=\pm e$. Then it is convenient to use a double unit cell with respect to that used in the Cooper pair box. This doubled cell has a reciprocal cell half the size of its Brillouin zone. The quasicharge within this new zone turns out to be a good quantum number: the allowed transitions are vertical, and a gap opens at the edges of this new Brillouin zone (see Fig.\ref{fig:5}b). It occurs because the degeneracy of the two unperturbed levels with momentum $\hbar k=\pm e/2$ is removed by the interaction with the environment. 

This causes a relevant change in the description of the Bloch oscillations. Indeed, in the presence of a weak current bias, the wave packet accelerates until it reaches the Brillouin zone boundary and jumps to the opposite border of the Brillouin zone: every time such umklapp event takes place, a single electron tunnels through the junction. 

The bias current turns out to be the sum of the current in the junction and the current in the ohmic resistor: it is the counterpart of the phonon-assisted transition in the spectra of the regular part of the charge response function. By increasing the current bias, the interband transitions cannot be neglected anymore, so that the Cooper pair current is restored. This explains the bump in the effective resistance observed experimentally at small current biases~\cite{Exp_2}. On the other hand, when the coupling increases, the gap between the first two effective bands with the lowest energy shrinks because of scattering with many phonons (the vertex of the interaction grows). At $\alpha>\alpha_c$, one expects that the gap is absent, and then the Bloch oscillations of polaronic Cooper pairs are restored. In fact, Cooper pairs are expected to be dressed by a large number of bare excitations of the baths. In the CL model, in principle, transfers of any quasicharge are allowed, so $\hbar k$, in the reduced Brillouin zone, is no longer a good quantum number. On the other hand, the plots in Fig.\ref{fig:4} point out that the main physical mechanisms are the same as the microscopic model, i.e. only the phonon-assisted transitions with $\hbar k=\pm e$ are activated, so that we expect that the previous scenario is not modified by changing the details of the interaction with the environment.      

\section{Conclusions}
We have proved that, in a small capacitance junction, the classical Josephson effect is not robust against quantum phase fluctuations and dissipative effects. When both of them are included in the Hamiltonian, a BKT quantum phase transition takes place, inducing localization of the phase particle at zero temperature. Within linear response theory, by using two different biasing schemes, the system turns out to be an insulator: the quantum phase fluctuations at low temperatures are proved to be ballistic and not diffusive. Consequences beyond the linear response regime are investigated in the presence of a current bias: we have predicted a transition from a dissipative quasiparticle current to a polaronic Cooper-pair current by increasing the coupling strength with the bosonic environment. 

A comment deserves a very large capacitance junction, i.e. transmon regime\cite{glazman1,glazman2,glazman3,glazman4}. In this case, characterized by $E_J/E_C\gg 1$, the bandwidth goes to zero exponentially, the phase particle mass can become very large, inducing a significant reduction of $D_Q$: it indicates that tunneling between two minima of the potential is a rare event. It is then clear that even a small temperature value will destroy the coherent motion of the phase particle, preventing the ballistic motion to set in and then, since the gap at $\hbar k=\pm \frac{e}{2}$ turns out to be vanishing, the system, in the presence of current bias, will exhibit only ideal conduction of Cooper pairs. 

\section{Acknowledgments}
G.D.F. acknowledges financial support from PNRR MUR Project No.
PE0000023-NQSTI. C.A.P. acknowledges funding from
IQARO (Spin-orbitronic Quantum Bits in Reconfigurable 2DOxides) project of the European Union’s Horizon Europe research and innovation programme under grant agreement n. 101115190. G.D.F.
and C.A.P. acknowledge funding from the PRIN 2022
project 2022FLSPAJ “Taming Noisy Quantum Dynamics” (TANQU). C.A.P. acknowledges funding from the
PRIN 2022 PNRR project P2022SB73K “Superconductivity in KTaO3 Oxide-2DEG NAnodevices for Topological quantum Applications” (SONATA) financed by the
European Union - Next Generation EU. The authors acknowledge illuminating discussions with professor R. Fazio.

\appendix

\section{The linear transport theory}{}\label{sec:app_transport}

In this Appendix, we generalize the concept of the response function to a generic observable $A$ and prove some useful relations employed in the main text. 

Given an observable $A$ and the operator $B$ so that $\frac{dB}{dt} =\frac{1}{i\hbar} [B,H] = A(t)$ and by defining $k = \frac{i}{\hbar}[A,B]$, the response function $\Psi_A(z)$ associated with the observable $A$ can be written as \cite{Mori}:
\begin{equation}\label{eqn:kubo_general1}
\Psi_A(z) = \frac{i}{z}\biggl[\Pi_A(z)+\langle k\rangle\biggr],   
\end{equation}
where $\Pi_A(z) = -\frac{i}{\hbar}\int_0^{+\infty}dt\,e^{izt}\bigl\langle [A(t),A(0)]\bigr\rangle$ is the Fourier transform of the correlation function associated with $A$. By taking the limit $\epsilon \to 0$ , the real part of the response function turns out to be\cite{Shastry_PRB, Evertz_PRB}:
\begin{equation}\label{eqn:kubo_tot1}
\Re\bigl[\Psi_A(\omega)\bigr] = D_A\delta(\omega) + \Psi_{A,\text{reg}}(\omega).
\end{equation}
Here, in the case $A=I$, even function $\Psi_{A,\text{reg}}(\omega)$ is the regular part of the conductivity  and $D_A$, the strength of the delta function, is the Drude weight or charge stiffness. 

By introducing the Matsubara Green function associated with the operator $A$,  $\Pi_A(\tau) = -\frac{1}{\hbar} \bigl\langle T_{\tau} A(\tau)A(0)\bigr\rangle $ ($\tau$ is imaginary time in the range $-\hbar\beta<\tau<\hbar\beta$ and $T_{\tau}$ is the time ordering operator), and the corresponding Fourier coefficients $\Pi_A(i\omega_n) = -\frac{1}{\hbar}\int_0^{\beta\hbar}d\tau \,\bigl\langle A(\tau)A(0)\bigr\rangle e^{i\omega_n \tau}$, $\omega_n$ being the Matsubara frequencies, it is possible to prove\cite{Evertz_PRB, Shastry_PRB}:

\begin{equation}\label{eqn:Drudea}
    D_A = \langle k \rangle + \Pi_{A}(i\omega_n \to 0).
\end{equation}
A property closely related to the Drude weight is the Meissner stiffness (it measures the superconducting density when $A$ is the current operator), defined by: 
\begin{equation}\label{eqn:Drudea1}
    D_{A,M} = \langle k \rangle + \Pi_{A}(i\omega_n =0).
\end{equation}
The difference between the two stiffnesses is a sum over all degenerate manifolds: 
\begin{equation}\label{eqn:Differenzaa}
    D_{A,M}-D_A = -\beta \sum_{E_n=E_m} p_n|\bigl \langle n| A | m \bigr \rangle|^2,
\end{equation}
$p_n$ being the Boltzmann weight of the eigenstate $|n\rangle$.  In the case of the admittance ($A=I$), the values of $D_I$ and $D_{I,M}$  allow to discriminate insulators, metals and superconductors.

It should be mentioned that there is an important relationship between the response function $\Psi_A(z)$ and the relaxation function introduced by Mori. Within the Mori formalism\cite{Mori} one defines an inner product between two operators $(A,B)=\
 \frac {1}{\beta \hbar} \int_{0}^{\beta \hbar} \left\langle e^{sH} A^{\dagger} e^{-sH} B\right\rangle ds$.  Then it is straightforward to prove that the function $\Psi_A(z)$ can be rewritten in terms of the Mori relaxation function $\Sigma_A(z)=\int_0^{+\infty}dt \,\Sigma_A(t)e^{izt}$:
\begin{equation}\label{eqn:mori_general}
\Psi_A(z) = \frac{i}{z} D_{A,M}+\beta \hbar (A,A) \Sigma_A(z),   
\end{equation}
where $\Sigma_A(t)=\frac{(A,A(t))}{(A,A)}$. This function describes the response of the system, in thermal equilibrium at $t=-\infty$, to a small classical field $h$ (which couples to the observable $A$) applied adiabatically from $t=-\infty$ and cut off at $t=0$. We emphasize that the study of the relaxation function associated with $A$ demands only the time evolution of $A$ under an equilibrium condition, independently of the applied field.  

Due to the crucial role played by the function $\Psi_A(z)$ in determining the response of the system to a small external disturbance, it is natural to ask which method allows one to obtain the dynamical spectra $\Psi_{A,\text{reg}}(\omega)$, the experimentally measurable quantity. At the end of this appendix, we will show that there is an exact relation between the function $\Psi_{A,\text{reg}}(\omega)$ and the function $\langle A(\tau)A(0)\rangle$: 
\begin{equation}\label{eqn:regular1}
    \int_{-\infty}^{+\infty}d\omega \,\frac{\hbar\omega}{2\pi}\Psi_{A,\text{reg}}(\omega)D_{\omega}(\tau) =\langle A(\tau)A(0)\rangle- \frac{D_{A} -D_{A,M}}{\beta},
\end{equation}
where $\tau$ ranges between $0$ and $\beta\hbar$. Then, by performing analytic continuation, for example, by using the maximum entropy method\cite{jarrell}, one can extract real frequency dynamical information from imaginary-time correlation functions computed in quantum Monte Carlo simulations.  The function $\Pi_A(\tau)$, coinciding with $-\frac{\langle A(\tau)A(0)\rangle}{\hbar}$ when $\tau\ge0$, allows us also to get $D_A$ and $D_{A,M}$ using Eq.\ref{eqn:Drudea} and Eq.\ref{eqn:Drudea1}, respectively\cite{Evertz_PRB}. In the main text, we applied the above described procedure to the two observables $A=I$ (in this case $B=-Q$) and $A=Q$ (here $B=C \phi$). 

Once the physical quantities $\Psi_{A,\text{reg}}(\omega)$, $D_A$  and $D_{A,M}$ are known, one can determine quantum fluctuations of the $B$ operator at real times: 
\begin{equation}\label{eqn:dynamic_general}
    \begin{split}
        &\frac{d}{dt}\biggl\langle\bigl(B(t)-B(0)\bigr)^2\biggr \rangle = \frac{2(D_A-D_{A,M})t}{\beta} \\&+\int_0^{+\infty} d\omega \, \frac{2\hbar\Psi_{A,\text{reg}}(\omega) \sin(\omega t)}{\pi\tanh(\beta\hbar\omega/2)},
    \end{split}
\end{equation}
i.e. the instantaneous charge and flux diffusivity for $A=I$ and $A=Q$, respectively.  

Now we prove Eq. \eqref{eqn:regular1} for $\tau>0$, starting from its left-hand side:
\begin{equation}
    \begin{split}\label{eq_app:regulara}
        \int_{-\infty}^{+\infty}d\omega\,\frac{\hbar\omega}{2\pi}\Psi_{A, \text{reg}}(\omega)\biggl[\frac{e^{-\omega\tau}}{1-e^{-\beta\hbar\omega}} 
+ \frac{e^{-\omega(\beta\hbar-\tau)}}{1-e^{-{\beta\hbar\omega}}}\biggr],
    \end{split}
\end{equation}
where we have used the fact that $D_\omega(\tau) = \cosh\bigl[\omega(\beta\hbar/2 -|\tau|)\bigr]/\sinh(\beta\hbar\omega/2)$. The integral in Eq. \eqref{eq_app:regulara} can be decomposed into the sum of two integrals:
\begin{subequations}\label{eq_app:regular_2_contributions}
    \begin{eqnarray}
        \int_{-\infty}^{+\infty}d\omega\,\frac{\hbar\omega}{2\pi}\Psi_{A, \text{reg}}(\omega)\frac{e^{-\omega\tau}}{1-e^{-\beta\hbar\omega}} \label{eq_app:regular_2_contributions_a} \\
        \int_{-\infty}^{+\infty}d\omega\,\frac{\hbar\omega}{2\pi}\Psi_{A, \text{reg}}(\omega)\frac{e^{-\omega(\beta\hbar-\tau)}}{1-e^{-\beta\hbar\omega}} \label{eq_app:regular_2_contributions_ba},
    \end{eqnarray}
\end{subequations}
that equally contribute, each yielding exactly half of the right-hand side of Eq.~\eqref{eqn:regular1}. In fact, the regular part of the response function can be expressed as a sum of delta functions peaked at each nonzero energy difference between the exact eigenstates of the Hamiltonian\cite{Shastry_PRB,Evertz_PRB}:
\begin{equation}\label{eq_app:decomposition_regulara}
    \begin{split}
        &\Psi_{A,\text{reg}}(\omega)=\\  &\pi\sum_{\substack{n,m \\ E_n\neq E_m}}\frac{\left|A_{nm}\right|^2}{E_m-E_n}\frac{e^{-\beta E_n} - e^{-\beta E_m}}{Z_p}\delta\biggl(\omega - \frac{E_m-E_n}{\hbar}\biggr),
    \end{split}
\end{equation}
where $E_n$ is the eigenvalue of $H$ associated with the eigenstate $|n\rangle$ and $A_{nm} = \langle n \left| A\right|m\rangle$ is the $(n,m)$ matrix element of the operator $A$ in the energy eigenstate basis. By substituting \eqref{eq_app:decomposition_regulara} into \eqref{eq_app:regular_2_contributions_a} we obtain:
\begin{equation}\label{eq_app:sum_n_neq_m}
    \begin{split}
        \frac{1}{2}\sum_{\substack{n,m\\E_n\neq E_m}} \left|A_{nm}\right|^2\frac{e^{-\beta E_n}}{Z_p}e^{-(E_m-E_n)\tau/\hbar}.
    \end{split}
\end{equation}
By adding and subtracting sum over all degenerate manifolds, we obtain:
\begin{equation}
    \begin{split}\label{eq_app:regular_sum_and_subtract}
        &\frac{1}{2}\sum_{n,m} \left|A_{nm}\right|^2\frac{e^{-\beta E_n}}{Z_p}e^{-(E_m-E_n)\tau/\hbar}\\&-\frac{1}{2}\sum_{\substack{n,m\\E_n=E_m}}|A_{nm}|^2\frac{e^{-\beta E_n}}{Z_p}.
    \end{split}
\end{equation}
The first term of Eq. \eqref{eq_app:regular_sum_and_subtract} corresponds to the decomposition of $\frac{\langle A(\tau)A(0)\rangle}{2}$ in the energy eigenstate basis, which exactly coincides with $-\frac{\hbar\Pi_A(\tau)}{2}$ for $\tau>0$. Using Eq. \eqref{eqn:Differenzaa}, it is straightforward to show that Eq. \eqref{eq_app:regular_sum_and_subtract} results in half of the right-hand side of \eqref{eqn:regular1}. The same approach can be applied to compute the integral in Eq. \ref{eq_app:regular_2_contributions_ba}. This procedure shows that we again obtain \eqref{eq_app:sum_n_neq_m}, thus proving Eq. \eqref{eqn:regular1} for $\tau>0$. 

Finally, we emphasize that, by merging Eq.\eqref{eq_app:decomposition_regulara} with Eq.\eqref{eqn:Drudea}, it is straightforward to demonstrate that the following sum rule is fulfilled: 
\begin{equation}
        \int_{-\infty}^{+\infty}d\omega\,\frac{\Psi_{A, \text{reg}}(\omega)}{\pi}=\langle k\rangle-D_A.
\end{equation}

\section{Mobility in the complex plane}
{}\label{sec:Mob}
In this Appendix, we give more details about the dimensionless mobility of the phase particle, defined by: $\mu(z)=\frac{1}{R}\frac{V(z)}{I(z)}$, where $R$ indicates the shunt resistance within the CL model ($R=R_S$) and the subgap resistance in the microscopic model ($R=R_{QP}$). Within the linear response theory, in the presence of a small current bias, the voltage across the junction can be expressed in terms of $\tilde{Q}(z)$, i.e. $V(z)=\frac{\tilde{Q}(z)}{C}$, and the relation between $\tilde{Q}(z)$ and $I(z)$ is given by Eq.\ref{eqn:qtz}, so the dimensionless mobility turns out to be: 
\begin{equation}\label{eqn:mob}
    \mu(z)=\frac{1}{R C^2}\frac{i}{z}\bigl(C+\Pi_Q(z)\bigr)=\frac{1}{R C^2}\Psi_Q(z),
\end{equation}
i.e. $\mu(z)$ is proportional to $\Psi_Q(z)$. 
Writing $\Pi_Q(z)=\Pi_Q(z)+\Pi_Q(i\omega_n\rightarrow 0)-\Pi_Q(i\omega_n\rightarrow 0)$, and using Eq.\ref{eqn:Drudea} with $A=Q$ and $k=C$, Eq.\ref{eqn:mob} can be recast in the following form: 
\begin{equation}\label{eqn:mob1}
    \mu(z)=\frac{1}{R C^2}\frac{i}{z} [D_Q+(\Pi_Q(z)-\Pi_Q(i\omega_n\rightarrow 0))].
\end{equation}
On the other hand, it is also possible to express the dimensionless mobility in terms of the Fourier transform of the phase-phase correlation function: 
\begin{equation}\label{eqn:corrphi}
\Pi_\varphi(z)=\int_0^{\infty}dt e^{i z t} \Pi_\varphi(t),
\end{equation}
where $\Pi_\varphi(t)=-\frac{i}{\hbar}\theta(t)\langle [\varphi(t),\varphi(0)]\rangle$. In fact, by performing a double integration by parts in Eq.\ref{eqn:corrphi}, it is straightforward to prove that: 
\begin{equation}\label{eqn:rel}
\Pi_\varphi(z)=(\frac{2\pi}{\phi_0 z})^2(\frac{1}{C}+\frac{1}{C^2}\Pi_Q(z)).
\end{equation}
Then, merging Eq.\ref{eqn:mob} and Eq.\ref{eqn:rel} yields:
\begin{equation}\label{eqn:relfin}
    \mu(z)=\frac{\alpha}{2\pi} i z \hbar \Pi_\varphi(z).
\end{equation}
We emphasize that all these relations are exact and represent different ways of expressing the dimensionless mobility of the phase particle in the complex plane. It is also clear that the physical quantity is the mobility evaluated on the real axis, i.e. $\mu(\omega)=\lim_{\epsilon \rightarrow 0^+}\mu(\omega+i\epsilon)$. Performing this limit in Eq.\ref{eqn:mob1} yields: 
\begin{equation}\label{eqn:mobreal}
    \mu(\omega)=\frac{1}{R C^2}(\pi D_Q \delta(\omega)+\Psi_{Q,reg}(\omega)).
\end{equation}
In particular, the dc mobility is given by $\mu_{dc}=\lim_{\omega\rightarrow 0^+}\mu(\omega)$, then it is related to the regular part of the charge response function $\Psi_{Q,reg}(\omega)$. The dc mobility determines the diffusive motion of the phase fluctuations in thermodynamic equilibrium, whereas the term containing $D_Q$ in Eq.\ref{eqn:mobreal} gives rise to the ballistic contribution.     

If $D_Q=0$, the dc mobility can be evaluated considering the limit $z\rightarrow 0$ in Eq.\ref{eqn:relfin}, performed along the imaginary axis. In fact, in this particular case, the limit is independent of the direction. On the other hand, if $D_Q\ne0$, the limit, performed along the imaginary axis, does not provide the dc mobility, since a singularity appears in the origin of the complex plane: it stems from the contribution proportional to $\frac{D_Q}{z}$ in Eq.\ref{eqn:mob1}. When $D_Q\ne0$, the dc mobility has to be calculated by first performing the analytic continuation on the real axis of $\mu(z)$, and, only then, carrying out the limit $\omega\rightarrow0$.

\begin{figure}
\flushleft
        \includegraphics[scale=0.34]{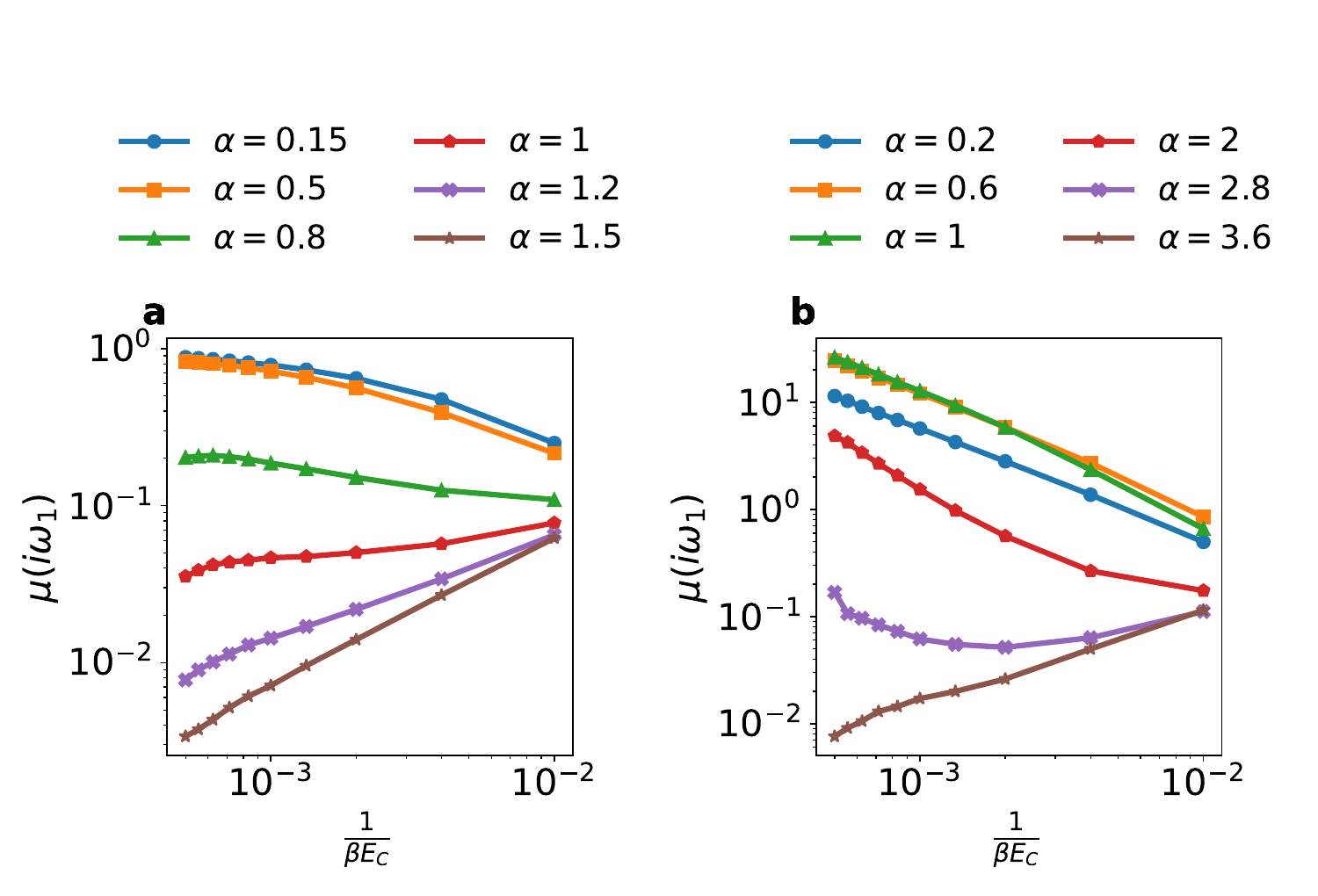}
        \caption{Mobility along the imaginary axis: plot of the dimensionless mobility $\mu(z)$ at the first Matsubara frequency $z = i\omega_1 = i\frac{2\pi}{\beta\hbar}$, for the CL model (a) and the QP model (b) for $E_J/E_C=0.5$. In (a), the curves tend to $1$ for $\alpha \to 0$ as $\omega_1$ approaches $0$, while they decay to $0$ for $\alpha>\alpha_c$. In (b), the mobility does not converge to $1$ for $\alpha \to 0$. } 
\label{fig:6}
\end{figure}

In Fig.\ref{fig:6} we plot $\mu(i\omega_n)$, with $n=1$, i.e. $\mu$ evaluated at the first Matsubara frequency, as a function of $\beta$ in both CL and QP models. As written in the main text, $\mu$ approaches $1$ when $\alpha \to 0$ and goes to $0$ for $\alpha>\alpha_c$ within the CL model (Fig.\ref{fig:6}a), as predicted in the literature\cite{Schmid}. In contrast, within the QP model, $\mu \neq 1$ for $\alpha \to 0$ (Fig.\ref{fig:6}b). However, we emphasize that, being $D_Q\ne0$ in both models, this limit does not represent the dc mobility at $T=0$. The plots in Fig.\ref{fig:3} and Fig.\ref{fig:4} point out that the dc mobility is zero indicating the absence of diffusive motion. Our results show that the phase particle motion, in a quantum Josephson junction, does not change from diffusive to localized, as is commonly believed, but from ballistic to localized.  

\section{The worldline Monte Carlo (WLMC) method}
{}\label{sec:MQ}

In this Appendix we explain the proposed worldline Monte Carlo approach in detail. We will specifically address the QP model, but the same Monte Carlo method can also be applied to the Caldeira-Leggett model. By using the path integral representation, the partition function associated with the QP model at $T=1/(K_B\beta)$ temperature is given by a sum over all periodic paths in imaginary time:
\begin{equation}\label{eqn:partiton_QP}
    Z_{QP} = \int_{\varphi(0) = \varphi(\beta\hbar)} D[\varphi(\tau)] \,e^{-\mathcal{S}_{QP}[\varphi(\tau)]/\hbar}
\end{equation}
Subsequently, any observable $A$ can be represented in this framework as $A[\varphi(\tau)]$ allowing us to evaluate its thermal equilibrium average:
\begin{equation}\label{eqn:mean_value_QP}
    \langle A\rangle = \frac{1}{Z_{QP}}\int_{\varphi(0) = \varphi(\beta\hbar)} D[\varphi(\tau)] \, A[\varphi(\tau)]e^{-\mathcal{S}_{QP}[\varphi(\tau)]/\hbar}
\end{equation}
The WLMC method is a Markov chain Monte Carlo based technique and it involves the use of the trotterization of the action:
\begin{equation}\label{eqn:disc_action}\begin{split}
    &\mathcal{S^{\text{dis}}_{QP}}[\{\varphi_m\}] = -E_J \Delta \tau \sum_{i=0}^{N-1}\cos\varphi_i+ \\
    &\sum_{i<j} K^{\text{dis}}_{ij}\sin^2\biggl(\frac{\varphi_i-\varphi_j}{4}\biggr)+\frac{\hbar^2}{4E_C \Delta \tau}\sum_{i=0}^{N-1}\bigl(\varphi_i - \varphi_{i+1}\bigr)^2
    \end{split}
\end{equation}
being $N = \frac{\beta}{\Delta \tau}$ and $K^{\text{dis}}_{ij} = \Delta \tau^2 K\bigl(\Delta\tau (i-j)\bigr)$. The discretization process consists of replacing the differential $d\tau$ with a finite difference $\Delta \tau$ and substituting continuous paths $\varphi(\tau)$ with discretized paths $\{\varphi_m\}$. In order to guarantee the periodicity of the paths, we define $\varphi_{N} = \varphi_0$. The integer $m \in [0, N-1]$ denotes an imaginary time step $0\le\tau_m = m\Delta \tau <\beta\hbar$. In this context, the partition function becomes:
\begin{equation}\label{eqn:disc_part}\begin{split}
    Z_{QP}^{\text{dis}} = \sum_{\{\varphi_m\}}e^{-\mathcal{S^{\text{dis}}_{QP}}[\{\varphi_m\}]/\hbar} 
    \end{split}.
\end{equation}
Equivalently, the functional integral \eqref{eqn:mean_value_QP} can be rewritten as:

\begin{equation}\label{eqn:disc_mean}\begin{split}
    \langle A^{\text{dis}} \rangle =\frac{1}{Z_{QP}^{\text{dis}}} \sum_{\{\varphi_m\}}A(\{\varphi_m\})e^{-\mathcal{S^{\text{dis}}_{QP}}[\{\varphi_m\}]/\hbar} 
    \end{split}.
\end{equation}
It is worth emphasizing that using \eqref{eqn:disc_part} and \eqref{eqn:disc_mean} instead of \eqref{eqn:partiton_QP} and \eqref{eqn:mean_value_QP} introduces an error of order $\Delta \tau$. The problem is now equivalent to a one-dimensional classical system of phase variables distributed on a chain of length $\beta$ and step $\Delta \tau$, interacting with each other. WLMC technique generate a Markov chain, which, after thermalization, samples the system's configurations $\{\varphi_m\}$, i.e. the worldlines, with the correct statistical weight, $p(\{\varphi_m\}) \propto e^{-\mathcal{S}_{QP}^{\text{dis}}[\{\varphi_m\}]/\hbar}$, simplifying the calculation of \eqref{eqn:disc_mean}. We adopted an efficient sampling of the paths, a variant of the cluster algorithm proposed by Werner and Troyer \cite{Troyer_MC,Troyer_PRL}, based on Wolff algorithm \cite{Wolff_PRL}. Starting from a worldline $\{\varphi^{\text{old}}_m\}$, we randomly select a root $j$ as the first site of the cluster, along with a symmetry axis $a \in [-L\pi, L\pi]$. The update move consists of a reflection move with respect to the axis $a$ of any site $l$ belonging to the cluster as:
\begin{equation}\label{eqn:update}\begin{split}
    \varphi_l^\text{new} = 2a -\varphi_l^{\text{old}} 
    \end{split}.
\end{equation}
We build up the cluster by connecting any site $u$ of the worldline to an already added site $v$ with Wolff like probability:
\begin{equation}\label{eqn:prob_wolff}\begin{split}
    P_W(u|v)  = \max\biggl\{0, 1 -e^{-[S_{R}(\varphi^{\text{old}}_u,\varphi_v^{\text{new}})-S_R(\varphi^{\text{new}}_u,\varphi_v^{\text{new}})]/\hbar} \biggr\}
    \end{split}
\end{equation}
where we have defined the retarded interaction as
\begin{equation}\label{eqn:ret_action}\begin{split}
    &S_R(\varphi_u,\varphi_v) =  K^{\text{dis}}_{ij}\sin^2\biggl(\frac{\varphi_u-\varphi_v}{4}\biggr) \\&+ \frac{\hbar^2}{4E_C\Delta \tau}(\varphi_u-\varphi_v)^2(\delta_{v,u+1} + \delta_{v,u-1}). 
    \end{split}
\end{equation}
Then we accept the reflection move of the whole cluster with a metropolis like probability:
\begin{equation}\label{eqn:prob_metr}\begin{split}
    &P_M\bigl(\{\varphi_m^{\text{new}}\} \to \{\varphi_m^{\text{old}}\}\bigl)  = \\
    &\min\biggl\{1,  e^{-[S_{J}(\{\varphi_m^{\text{new}}\})-S_{J}(\{\varphi_m^{\text{old}}\})]/\hbar} \biggr\}
    \end{split}
\end{equation}
where $S_J\bigl(\{\varphi_m\}\bigr) = -E_J\Delta\tau\sum_{i=0}^{N-1}\cos\varphi_i$ is the static interaction along the worldline. The algorithm we exploited is made up of two micromoves. The first one require the axis to be chosen among the symmetry points of the Josephson potential: $a = k\pi$, where $k$ is an integer in the range $[-L, L]$ as proposed in the original work by Werner and Troyer \cite{Troyer_MC, Troyer_PRL}. Such choice of the axis ensure the acceptance probability \eqref{eqn:prob_metr} to be $1$. A more efficient option is to choose the axis among the three multiples of $\pi$ closest to the root. The use of this micromove alone does not guarantee the ergodicity of the algorithm, therefore we perform a second update where the axis is chosen randomly in the neighbourhood of the root. Such micromove has a nonzero probability of generating single-site clusters. Therefore, since the reflection axes are randomly chosen, it ensures ergodicity, as every system configuration can be reached within a finite number of moves. The parameter $L$ should be large enough to contain the entire worldline, and it can be determined during the thermalization process. It is worth to stress that even $L$ being finite it will be sufficiently large that the boundaries no longer affect the system: this procedure is then equivalent to consider the extended picture of the phase variable. An important aspect to consider is that we define a Monte Carlo step as a sequence of micromoves, both with a random axis and located at potential minima or maxima, that on average attempt to update $N$ sites, i.e. the entire worldline.. Because of the nature of the long-range interaction, the time for a complete Monte Carlo step becomes $O(N^2)$, thus preventing an efficient exploration of large values of $\beta$. To address this issue, the algorithm can be further improved by incorporating the approach proposed by Luijten and Bl\"{o}te \cite{Luijten1, Luijten2}. First, we define a function $F_{\left|u-v\right|} $ such that
\begin{equation}
    \begin{split}
        F_{\left|u-v\right|} \ge S_R(\varphi_u^{\text{old}},\varphi_u^{\text{new}}) - S_R(\varphi_u^{\text{new}},\varphi_u^{\text{new}}) 
    \end{split}
\end{equation}
which depends only on $u$ and $v$, independently of $\varphi_u$ and $\varphi_v$. The cluster growth process is divided into two stages. The first stage involves the creation of provisional bonds with probability given by 
\begin{equation}\label{eqn:luijten_provv}\begin{split}
    P_{1LB}(u|v)  = 1 -e^{-F_{\left|u-v\right|}/\hbar}. 
    \end{split}
\end{equation}
This procedure can be implemented as follows: starting from a site $u$ we define the probability of forming no bonds from site $u+1$ up to $u+k-1$ and creating a bond between $u$ and $k$ as 
\begin{equation}\label{eqn:luijten_provv_up_to_k}\begin{split}
    P(k) = e^{-(F_1+F_2+\dots+F_{k-1})/\hbar}\biggl(1-e^{-F_{k}/\hbar}\biggr).
    \end{split}
\end{equation}
The associated cumulative probability is given by:
\begin{equation}\label{eqn:luijten_cum}\begin{split}
    C(k) = \sum_{k'=1}^{k} P(k')
    \end{split}
\end{equation}
No bonds will be formed with probability $1-C(N)$, while the first bond between $u$ and $u+k_1$ will be created with probability $C(k_1)-C(k_1-1)$. In the latter case we define the probability of forming no bonds from $u+k_1+1$ up to $u+k-1$ while creating a bond between $u$ and $k$ as
\begin{equation*}\begin{split}
    P_2(k) &= e^{-(F_{k_1+1}+F_{k_1+2}+\dots+F_{k-1})/\hbar}\biggl(1-e^{-F_{k}/\hbar}\biggr) \\&= \frac{P(k)}{1-C(k_1)}.
    \end{split}
\end{equation*}
The corresponding cumulative probability is:
\begin{equation}\begin{split}
    C_2(k) = \sum_{k'=k_1+1}^{k} P(k') = \frac{C(k)-C(k_1)}{1-C(k_1)}.
    \end{split}
\end{equation}
No further bonds will be formed with probability $1-C_2(N)$, while the second bond between $u$ and $u+k_2$ will be created with probability $C_2(k_2)-C_2(k_2-1)$. This process continues, iteratively connecting sites to $u$, until no more bonds are created. The same scheme is then repeated for each site added to the provisional cluster. The primary advantage of this approach is that, by employing a bisection algorithm, the bond selection per site can be performed in $O(\log N)$ time instead of $O(N)$ since $F_{\left|u-v\right|}$ is independent of the specific configurations of $u$ and $v$. In the second stage, each provisional bond is confirmed with probability
\begin{equation*}
\begin{split}
    &P_{2LB}(u|v)  = \\ &\frac{\max \biggl\{0, 1-e^{-[S_{R}(\varphi^{\text{old}}_u,\varphi_v^{\text{new}})-S_R(\varphi^{\text{new}}_u,\varphi_v^{\text{new}})]/\hbar}\biggr\}}{1 -e^{-F_{\left|u-v\right|}/\hbar}}. 
    \end{split}
\end{equation*}


\newpage

\bibliography{main_text}

\begin{thebibliography}{60}%
\makeatletter
\providecommand \@ifxundefined [1]{%
 \@ifx{#1\undefined}
}%
\providecommand \@ifnum [1]{%
 \ifnum #1\expandafter \@firstoftwo
 \else \expandafter \@secondoftwo
 \fi
}%
\providecommand \@ifx [1]{%
 \ifx #1\expandafter \@firstoftwo
 \else \expandafter \@secondoftwo
 \fi
}%
\providecommand \natexlab [1]{#1}%
\providecommand \enquote  [1]{``#1''}%
\providecommand \bibnamefont  [1]{#1}%
\providecommand \bibfnamefont [1]{#1}%
\providecommand \citenamefont [1]{#1}%
\providecommand \href@noop [0]{\@secondoftwo}%
\providecommand \href [0]{\begingroup \@sanitize@url \@href}%
\providecommand \@href[1]{\@@startlink{#1}\@@href}%
\providecommand \@@href[1]{\endgroup#1\@@endlink}%
\providecommand \@sanitize@url [0]{\catcode `\\12\catcode `\$12\catcode `\&12\catcode `\#12\catcode `\^12\catcode `\_12\catcode `\%12\relax}%
\providecommand \@@startlink[1]{}%
\providecommand \@@endlink[0]{}%
\providecommand \url  [0]{\begingroup\@sanitize@url \@url }%
\providecommand \@url [1]{\endgroup\@href {#1}{\urlprefix }}%
\providecommand \urlprefix  [0]{URL }%
\providecommand \Eprint [0]{\href }%
\providecommand \doibase [0]{https://doi.org/}%
\providecommand \selectlanguage [0]{\@gobble}%
\providecommand \bibinfo  [0]{\@secondoftwo}%
\providecommand \bibfield  [0]{\@secondoftwo}%
\providecommand \translation [1]{[#1]}%
\providecommand \BibitemOpen [0]{}%
\providecommand \bibitemStop [0]{}%
\providecommand \bibitemNoStop [0]{.\EOS\space}%
\providecommand \EOS [0]{\spacefactor3000\relax}%
\providecommand \BibitemShut  [1]{\csname bibitem#1\endcsname}%
\let\auto@bib@innerbib\@empty
\bibitem [{\citenamefont {Weiss}(1999)}]{Weiss1999}%
  \BibitemOpen
  \bibfield  {author} {\bibinfo {author} {\bibfnamefont {U.}~\bibnamefont {Weiss}},\ }\href@noop {} {\emph {\bibinfo {title} {Quantum Dissipative Systems}}},\ \bibinfo {edition} {2nd}\ ed.,\ \bibinfo {series} {Series in Modern Condensed Matter Physics}, Vol.~\bibinfo {volume} {10}\ (\bibinfo  {publisher} {World Scientific},\ \bibinfo {year} {1999})\BibitemShut {NoStop}%
\bibitem [{\citenamefont {Nitzan}(2006)}]{nitzan_chemical}%
  \BibitemOpen
  \bibfield  {author} {\bibinfo {author} {\bibfnamefont {A.}~\bibnamefont {Nitzan}},\ }\href@noop {} {\emph {\bibinfo {title} {Chemical Dynamics in Condensed Phases: Relaxation, Transfer, and Reactions in Condensed Molecular Systems}}}\ (\bibinfo  {publisher} {Oxford University Press},\ \bibinfo {address} {New York},\ \bibinfo {year} {2006})\BibitemShut {NoStop}%
\bibitem [{\citenamefont {Leggett}\ \emph {et~al.}(1987)\citenamefont {Leggett}, \citenamefont {Chakravarty}, \citenamefont {Dorsey}, \citenamefont {Fisher}, \citenamefont {Garg},\ and\ \citenamefont {Zwerger}}]{Leggett_spinboson}%
  \BibitemOpen
  \bibfield  {author} {\bibinfo {author} {\bibfnamefont {A.~J.}\ \bibnamefont {Leggett}}, \bibinfo {author} {\bibfnamefont {S.}~\bibnamefont {Chakravarty}}, \bibinfo {author} {\bibfnamefont {A.~T.}\ \bibnamefont {Dorsey}}, \bibinfo {author} {\bibfnamefont {M.~P.~A.}\ \bibnamefont {Fisher}}, \bibinfo {author} {\bibfnamefont {A.}~\bibnamefont {Garg}},\ and\ \bibinfo {author} {\bibfnamefont {W.}~\bibnamefont {Zwerger}},\ }\href {https://doi.org/10.1103/RevModPhys.59.1} {\bibfield  {journal} {\bibinfo  {journal} {Rev. Mod. Phys.}\ }\textbf {\bibinfo {volume} {59}},\ \bibinfo {pages} {1} (\bibinfo {year} {1987})}\BibitemShut {NoStop}%
\bibitem [{\citenamefont {May}\ and\ \citenamefont {Kuhn}(2004)}]{may_molecular_systems}%
  \BibitemOpen
  \bibfield  {author} {\bibinfo {author} {\bibfnamefont {V.}~\bibnamefont {May}}\ and\ \bibinfo {author} {\bibfnamefont {O.}~\bibnamefont {Kuhn}},\ }\href@noop {} {\emph {\bibinfo {title} {Charge and Energy Transfer Dynamics in Molecular Systems}}}\ (\bibinfo  {publisher} {Wiley-VCH},\ \bibinfo {address} {Weinheim},\ \bibinfo {year} {2004})\BibitemShut {NoStop}%
\bibitem [{\citenamefont {Nielsen}\ and\ \citenamefont {Chuang}(2000)}]{nielsen_qc_qinfo}%
  \BibitemOpen
  \bibfield  {author} {\bibinfo {author} {\bibfnamefont {M.~A.}\ \bibnamefont {Nielsen}}\ and\ \bibinfo {author} {\bibfnamefont {I.~L.}\ \bibnamefont {Chuang}},\ }\href@noop {} {\emph {\bibinfo {title} {Quantum Computation and Quantum Information}}}\ (\bibinfo  {publisher} {Cambridge University Press},\ \bibinfo {address} {Cambridge},\ \bibinfo {year} {2000})\BibitemShut {NoStop}%
\bibitem [{\citenamefont {Alipour}\ \emph {et~al.}(2014)\citenamefont {Alipour}, \citenamefont {Mehboudi},\ and\ \citenamefont {Rezakhani}}]{Alipour_metrology}%
  \BibitemOpen
  \bibfield  {author} {\bibinfo {author} {\bibfnamefont {S.}~\bibnamefont {Alipour}}, \bibinfo {author} {\bibfnamefont {M.}~\bibnamefont {Mehboudi}},\ and\ \bibinfo {author} {\bibfnamefont {A.~T.}\ \bibnamefont {Rezakhani}},\ }\href {https://doi.org/10.1103/PhysRevLett.112.120405} {\bibfield  {journal} {\bibinfo  {journal} {Phys. Rev. Lett.}\ }\textbf {\bibinfo {volume} {112}},\ \bibinfo {pages} {120405} (\bibinfo {year} {2014})}\BibitemShut {NoStop}%
\bibitem [{\citenamefont {Gardiner}\ and\ \citenamefont {Zoller}(2004)}]{gardiner_zoller_QuantumNoise}%
  \BibitemOpen
  \bibfield  {author} {\bibinfo {author} {\bibfnamefont {C.~W.}\ \bibnamefont {Gardiner}}\ and\ \bibinfo {author} {\bibfnamefont {P.}~\bibnamefont {Zoller}},\ }\href@noop {} {\emph {\bibinfo {title} {Quantum Noise: A Handbook of Markovian and Non-Markovian Quantum Stochastic Methods with Applications to Quantum Optics}}},\ \bibinfo {edition} {3rd}\ ed.\ (\bibinfo  {publisher} {Springer},\ \bibinfo {address} {Berlin, Heidelberg},\ \bibinfo {year} {2004})\BibitemShut {NoStop}%
\bibitem [{\citenamefont {Lidar}(2019)}]{Lidar}%
  \BibitemOpen
  \bibfield  {author} {\bibinfo {author} {\bibfnamefont {D.~A.}\ \bibnamefont {Lidar}},\ }\href@noop {} {\  (\bibinfo {year} {2019})},\ \Eprint {https://arxiv.org/abs/1902.00967} {arXiv:1902.00967 [quant-ph]} \BibitemShut {NoStop}%
\bibitem [{\citenamefont {Breuer}\ \emph {et~al.}(2016)\citenamefont {Breuer}, \citenamefont {Laine}, \citenamefont {Piilo},\ and\ \citenamefont {Vacchini}}]{Breuer_RevModPhys_nonmarkovian}%
  \BibitemOpen
  \bibfield  {author} {\bibinfo {author} {\bibfnamefont {H.-P.}\ \bibnamefont {Breuer}}, \bibinfo {author} {\bibfnamefont {E.-M.}\ \bibnamefont {Laine}}, \bibinfo {author} {\bibfnamefont {J.}~\bibnamefont {Piilo}},\ and\ \bibinfo {author} {\bibfnamefont {B.}~\bibnamefont {Vacchini}},\ }\href {https://doi.org/10.1103/RevModPhys.88.021002} {\bibfield  {journal} {\bibinfo  {journal} {Rev. Mod. Phys.}\ }\textbf {\bibinfo {volume} {88}},\ \bibinfo {pages} {021002} (\bibinfo {year} {2016})}\BibitemShut {NoStop}%
\bibitem [{\citenamefont {Plenio}\ \emph {et~al.}(2014)\citenamefont {Plenio}, \citenamefont {Huelga},\ and\ \citenamefont {Rivas}}]{Rivas_nonmarkvovian}%
  \BibitemOpen
  \bibfield  {author} {\bibinfo {author} {\bibfnamefont {M.~B.}\ \bibnamefont {Plenio}}, \bibinfo {author} {\bibfnamefont {S.~F.}\ \bibnamefont {Huelga}},\ and\ \bibinfo {author} {\bibfnamefont {A.}~\bibnamefont {Rivas}},\ }\href {https://doi.org/10.1088/0034-4885/77/9/094001} {\bibfield  {journal} {\bibinfo  {journal} {Rept. Prog. Phys.}\ }\textbf {\bibinfo {volume} {77}},\ \bibinfo {pages} {094001} (\bibinfo {year} {2014})}\BibitemShut {NoStop}%
\bibitem [{\citenamefont {Chru\ifmmode \acute{s}\else \'{s}\fi{}ci\ifmmode~\acute{n}\else \'{n}\fi{}ski}\ and\ \citenamefont {Maniscalco}(2014)}]{Maniscalco_nonmarkovian}%
  \BibitemOpen
  \bibfield  {author} {\bibinfo {author} {\bibfnamefont {D.}~\bibnamefont {Chru\ifmmode \acute{s}\else \'{s}\fi{}ci\ifmmode~\acute{n}\else \'{n}\fi{}ski}}\ and\ \bibinfo {author} {\bibfnamefont {S.}~\bibnamefont {Maniscalco}},\ }\href {https://doi.org/10.1103/PhysRevLett.112.120404} {\bibfield  {journal} {\bibinfo  {journal} {Phys. Rev. Lett.}\ }\textbf {\bibinfo {volume} {112}},\ \bibinfo {pages} {120404} (\bibinfo {year} {2014})}\BibitemShut {NoStop}%
\bibitem [{\citenamefont {Rabi}(1936)}]{Rabi_original}%
  \BibitemOpen
  \bibfield  {author} {\bibinfo {author} {\bibfnamefont {I.~I.}\ \bibnamefont {Rabi}},\ }\href {https://doi.org/10.1103/PhysRev.49.324} {\bibfield  {journal} {\bibinfo  {journal} {Phys. Rev.}\ }\textbf {\bibinfo {volume} {49}},\ \bibinfo {pages} {324} (\bibinfo {year} {1936})}\BibitemShut {NoStop}%
\bibitem [{\citenamefont {Jaynes}\ and\ \citenamefont {Cummings}(1963)}]{JCM_rabi}%
  \BibitemOpen
  \bibfield  {author} {\bibinfo {author} {\bibfnamefont {E.}~\bibnamefont {Jaynes}}\ and\ \bibinfo {author} {\bibfnamefont {F.}~\bibnamefont {Cummings}},\ }\href {https://doi.org/10.1109/PROC.1963.1664} {\bibfield  {journal} {\bibinfo  {journal} {Proceedings of the IEEE}\ }\textbf {\bibinfo {volume} {51}},\ \bibinfo {pages} {89} (\bibinfo {year} {1963})}\BibitemShut {NoStop}%
\bibitem [{\citenamefont {Miller}\ \emph {et~al.}(2005)\citenamefont {Miller}, \citenamefont {Northup}, \citenamefont {Birnbaum}, \citenamefont {Boca}, \citenamefont {Boozer},\ and\ \citenamefont {Kimble}}]{Rabi_optics1}%
  \BibitemOpen
  \bibfield  {author} {\bibinfo {author} {\bibfnamefont {R.}~\bibnamefont {Miller}}, \bibinfo {author} {\bibfnamefont {T.~E.}\ \bibnamefont {Northup}}, \bibinfo {author} {\bibfnamefont {K.~M.}\ \bibnamefont {Birnbaum}}, \bibinfo {author} {\bibfnamefont {A.}~\bibnamefont {Boca}}, \bibinfo {author} {\bibfnamefont {A.~D.}\ \bibnamefont {Boozer}},\ and\ \bibinfo {author} {\bibfnamefont {H.~J.}\ \bibnamefont {Kimble}},\ }\href {https://doi.org/10.1088/0953-4075/38/9/007} {\bibfield  {journal} {\bibinfo  {journal} {J. Phys. B: At. Mol. Opt. Phys}\ }\textbf {\bibinfo {volume} {38}},\ \bibinfo {pages} {S551} (\bibinfo {year} {2005})}\BibitemShut {NoStop}%
\bibitem [{\citenamefont {Walther}\ \emph {et~al.}(2006)\citenamefont {Walther}, \citenamefont {Varcoe}, \citenamefont {Englert},\ and\ \citenamefont {Becker}}]{Rabi_optics2}%
  \BibitemOpen
  \bibfield  {author} {\bibinfo {author} {\bibfnamefont {H.}~\bibnamefont {Walther}}, \bibinfo {author} {\bibfnamefont {B.~T.~H.}\ \bibnamefont {Varcoe}}, \bibinfo {author} {\bibfnamefont {B.-G.}\ \bibnamefont {Englert}},\ and\ \bibinfo {author} {\bibfnamefont {T.}~\bibnamefont {Becker}},\ }\href {https://doi.org/10.1088/0034-4885/69/5/r02} {\bibfield  {journal} {\bibinfo  {journal} {Rept. Prog. Phys.}\ }\textbf {\bibinfo {volume} {69}},\ \bibinfo {pages} {1325–1382} (\bibinfo {year} {2006})}\BibitemShut {NoStop}%
\bibitem [{\citenamefont {Haroche}\ and\ \citenamefont {Raimond}(2006)}]{Rabi_optics3}%
  \BibitemOpen
  \bibfield  {author} {\bibinfo {author} {\bibfnamefont {S.}~\bibnamefont {Haroche}}\ and\ \bibinfo {author} {\bibfnamefont {J.-M.}\ \bibnamefont {Raimond}},\ }\href@noop {} {\emph {\bibinfo {title} {Exploring the Quantum: Atoms, Cavities, and Photons}}}\ (\bibinfo  {publisher} {Oxford University Press},\ \bibinfo {year} {2006})\BibitemShut {NoStop}%
\bibitem [{\citenamefont {De~Filippis}\ \emph {et~al.}(2020)\citenamefont {De~Filippis}, \citenamefont {de~Candia}, \citenamefont {Cangemi}, \citenamefont {Sassetti}, \citenamefont {Fazio},\ and\ \citenamefont {Cataudella}}]{Giulio_spinboson_PRB}%
  \BibitemOpen
  \bibfield  {author} {\bibinfo {author} {\bibfnamefont {G.}~\bibnamefont {De~Filippis}}, \bibinfo {author} {\bibfnamefont {A.}~\bibnamefont {de~Candia}}, \bibinfo {author} {\bibfnamefont {L.~M.}\ \bibnamefont {Cangemi}}, \bibinfo {author} {\bibfnamefont {M.}~\bibnamefont {Sassetti}}, \bibinfo {author} {\bibfnamefont {R.}~\bibnamefont {Fazio}},\ and\ \bibinfo {author} {\bibfnamefont {V.}~\bibnamefont {Cataudella}},\ }\href {https://doi.org/10.1103/PhysRevB.101.180408} {\bibfield  {journal} {\bibinfo  {journal} {Phys. Rev. B}\ }\textbf {\bibinfo {volume} {101}},\ \bibinfo {pages} {180408} (\bibinfo {year} {2020})}\BibitemShut {NoStop}%
\bibitem [{\citenamefont {De~Filippis}\ \emph {et~al.}(2021)\citenamefont {De~Filippis}, \citenamefont {de~Candia}, \citenamefont {Mishchenko}, \citenamefont {Cangemi}, \citenamefont {Nocera}, \citenamefont {Mishchenko}, \citenamefont {Sassetti}, \citenamefont {Fazio}, \citenamefont {Nagaosa},\ and\ \citenamefont {Cataudella}}]{Giulio_manyspin_PRB}%
  \BibitemOpen
  \bibfield  {author} {\bibinfo {author} {\bibfnamefont {G.}~\bibnamefont {De~Filippis}}, \bibinfo {author} {\bibfnamefont {A.}~\bibnamefont {de~Candia}}, \bibinfo {author} {\bibfnamefont {A.~S.}\ \bibnamefont {Mishchenko}}, \bibinfo {author} {\bibfnamefont {L.~M.}\ \bibnamefont {Cangemi}}, \bibinfo {author} {\bibfnamefont {A.}~\bibnamefont {Nocera}}, \bibinfo {author} {\bibfnamefont {P.~A.}\ \bibnamefont {Mishchenko}}, \bibinfo {author} {\bibfnamefont {M.}~\bibnamefont {Sassetti}}, \bibinfo {author} {\bibfnamefont {R.}~\bibnamefont {Fazio}}, \bibinfo {author} {\bibfnamefont {N.}~\bibnamefont {Nagaosa}},\ and\ \bibinfo {author} {\bibfnamefont {V.}~\bibnamefont {Cataudella}},\ }\href {https://doi.org/10.1103/PhysRevB.104.L060410} {\bibfield  {journal} {\bibinfo  {journal} {Phys. Rev. B}\ }\textbf {\bibinfo {volume} {104}},\ \bibinfo {pages} {L060410} (\bibinfo {year} {2021})}\BibitemShut {NoStop}%
\bibitem [{\citenamefont {Di~Bello}\ \emph {et~al.}(2024)\citenamefont {Di~Bello}, \citenamefont {Ponticelli}, \citenamefont {Pavan}, \citenamefont {Cataudella}, \citenamefont {De~Filippis}, \citenamefont {de~Candia},\ and\ \citenamefont {Perroni}}]{Grazia_NatComm}%
  \BibitemOpen
  \bibfield  {author} {\bibinfo {author} {\bibfnamefont {G.}~\bibnamefont {Di~Bello}}, \bibinfo {author} {\bibfnamefont {A.}~\bibnamefont {Ponticelli}}, \bibinfo {author} {\bibfnamefont {F.}~\bibnamefont {Pavan}}, \bibinfo {author} {\bibfnamefont {V.}~\bibnamefont {Cataudella}}, \bibinfo {author} {\bibfnamefont {G.}~\bibnamefont {De~Filippis}}, \bibinfo {author} {\bibfnamefont {A.}~\bibnamefont {de~Candia}},\ and\ \bibinfo {author} {\bibfnamefont {C.~A.}\ \bibnamefont {Perroni}},\ }\href {https://doi.org/10.1038/s42005-024-01855-8} {\bibfield  {journal} {\bibinfo  {journal} {Commun. Phys.}\ }\textbf {\bibinfo {volume} {7}},\ \bibinfo {pages} {364} (\bibinfo {year} {2024})}\BibitemShut {NoStop}%
\bibitem [{\citenamefont {De~Filippis}\ \emph {et~al.}(2023)\citenamefont {De~Filippis}, \citenamefont {de~Candia}, \citenamefont {Di~Bello}, \citenamefont {Perroni}, \citenamefont {Cangemi}, \citenamefont {Nocera}, \citenamefont {Sassetti}, \citenamefont {Fazio},\ and\ \citenamefont {Cataudella}}]{Giulio_PRL}%
  \BibitemOpen
  \bibfield  {author} {\bibinfo {author} {\bibfnamefont {G.}~\bibnamefont {De~Filippis}}, \bibinfo {author} {\bibfnamefont {A.}~\bibnamefont {de~Candia}}, \bibinfo {author} {\bibfnamefont {G.}~\bibnamefont {Di~Bello}}, \bibinfo {author} {\bibfnamefont {C.~A.}\ \bibnamefont {Perroni}}, \bibinfo {author} {\bibfnamefont {L.~M.}\ \bibnamefont {Cangemi}}, \bibinfo {author} {\bibfnamefont {A.}~\bibnamefont {Nocera}}, \bibinfo {author} {\bibfnamefont {M.}~\bibnamefont {Sassetti}}, \bibinfo {author} {\bibfnamefont {R.}~\bibnamefont {Fazio}},\ and\ \bibinfo {author} {\bibfnamefont {V.}~\bibnamefont {Cataudella}},\ }\href {https://doi.org/10.1103/PhysRevLett.130.210404} {\bibfield  {journal} {\bibinfo  {journal} {Phys. Rev. Lett.}\ }\textbf {\bibinfo {volume} {130}},\ \bibinfo {pages} {210404} (\bibinfo {year} {2023})}\BibitemShut {NoStop}%
\bibitem [{\citenamefont {Murani}\ \emph {et~al.}(2020)\citenamefont {Murani}, \citenamefont {Bourlet}, \citenamefont {le~Sueur}, \citenamefont {Portier}, \citenamefont {Altimiras}, \citenamefont {Esteve}, \citenamefont {Grabert}, \citenamefont {Stockburger}, \citenamefont {Ankerhold},\ and\ \citenamefont {Joyez}}]{Murani_absence}%
  \BibitemOpen
  \bibfield  {author} {\bibinfo {author} {\bibfnamefont {A.}~\bibnamefont {Murani}}, \bibinfo {author} {\bibfnamefont {N.}~\bibnamefont {Bourlet}}, \bibinfo {author} {\bibfnamefont {H.}~\bibnamefont {le~Sueur}}, \bibinfo {author} {\bibfnamefont {F.}~\bibnamefont {Portier}}, \bibinfo {author} {\bibfnamefont {C.}~\bibnamefont {Altimiras}}, \bibinfo {author} {\bibfnamefont {D.}~\bibnamefont {Esteve}}, \bibinfo {author} {\bibfnamefont {H.}~\bibnamefont {Grabert}}, \bibinfo {author} {\bibfnamefont {J.}~\bibnamefont {Stockburger}}, \bibinfo {author} {\bibfnamefont {J.}~\bibnamefont {Ankerhold}},\ and\ \bibinfo {author} {\bibfnamefont {P.}~\bibnamefont {Joyez}},\ }\href {https://doi.org/10.1103/PhysRevX.10.021003} {\bibfield  {journal} {\bibinfo  {journal} {Phys. Rev. X}\ }\textbf {\bibinfo {volume} {10}},\ \bibinfo {pages} {021003} (\bibinfo {year} {2020})}\BibitemShut {NoStop}%
\bibitem [{\citenamefont {Hakonen}\ and\ \citenamefont {Sonin}(2021)}]{Comment_on_absence}%
  \BibitemOpen
  \bibfield  {author} {\bibinfo {author} {\bibfnamefont {P.~J.}\ \bibnamefont {Hakonen}}\ and\ \bibinfo {author} {\bibfnamefont {E.~B.}\ \bibnamefont {Sonin}},\ }\href {https://doi.org/10.1103/PhysRevX.11.018001} {\bibfield  {journal} {\bibinfo  {journal} {Phys. Rev. X}\ }\textbf {\bibinfo {volume} {11}},\ \bibinfo {pages} {018001} (\bibinfo {year} {2021})}\BibitemShut {NoStop}%
\bibitem [{\citenamefont {Murani}\ \emph {et~al.}(2021)\citenamefont {Murani}, \citenamefont {Bourlet}, \citenamefont {le~Sueur}, \citenamefont {Portier}, \citenamefont {Altimiras}, \citenamefont {Esteve}, \citenamefont {Grabert}, \citenamefont {Stockburger}, \citenamefont {Ankerhold},\ and\ \citenamefont {Joyez}}]{Reply_to_comment}%
  \BibitemOpen
  \bibfield  {author} {\bibinfo {author} {\bibfnamefont {A.}~\bibnamefont {Murani}}, \bibinfo {author} {\bibfnamefont {N.}~\bibnamefont {Bourlet}}, \bibinfo {author} {\bibfnamefont {H.}~\bibnamefont {le~Sueur}}, \bibinfo {author} {\bibfnamefont {F.}~\bibnamefont {Portier}}, \bibinfo {author} {\bibfnamefont {C.}~\bibnamefont {Altimiras}}, \bibinfo {author} {\bibfnamefont {D.}~\bibnamefont {Esteve}}, \bibinfo {author} {\bibfnamefont {H.}~\bibnamefont {Grabert}}, \bibinfo {author} {\bibfnamefont {J.}~\bibnamefont {Stockburger}}, \bibinfo {author} {\bibfnamefont {J.}~\bibnamefont {Ankerhold}},\ and\ \bibinfo {author} {\bibfnamefont {P.}~\bibnamefont {Joyez}},\ }\href {https://doi.org/10.1103/PhysRevX.11.018002} {\bibfield  {journal} {\bibinfo  {journal} {Phys. Rev. X}\ }\textbf {\bibinfo {volume} {11}},\ \bibinfo {pages} {018002} (\bibinfo {year} {2021})}\BibitemShut {NoStop}%
\bibitem [{\citenamefont {Schmid}(1983)}]{Schmid}%
  \BibitemOpen
  \bibfield  {author} {\bibinfo {author} {\bibfnamefont {A.}~\bibnamefont {Schmid}},\ }\href {https://doi.org/10.1103/PhysRevLett.51.1506} {\bibfield  {journal} {\bibinfo  {journal} {Phys. Rev. Lett.}\ }\textbf {\bibinfo {volume} {51}},\ \bibinfo {pages} {1506} (\bibinfo {year} {1983})}\BibitemShut {NoStop}%
\bibitem [{\citenamefont {Bulgadaev}(1984)}]{Bulgadaev}%
  \BibitemOpen
  \bibfield  {author} {\bibinfo {author} {\bibfnamefont {S.}~\bibnamefont {Bulgadaev}},\ }\href@noop {} {\bibfield  {journal} {\bibinfo  {journal} {ZhETF Pisma Redaktsiiu}\ } (\bibinfo {year} {1984})}\BibitemShut {NoStop}%
\bibitem [{\citenamefont {Yagi}\ \emph {et~al.}(1997)\citenamefont {Yagi}, \citenamefont {Kobayashi},\ and\ \citenamefont {Ootuka}}]{Exp_1}%
  \BibitemOpen
  \bibfield  {author} {\bibinfo {author} {\bibfnamefont {R.}~\bibnamefont {Yagi}}, \bibinfo {author} {\bibfnamefont {S.-i.}\ \bibnamefont {Kobayashi}},\ and\ \bibinfo {author} {\bibfnamefont {Y.}~\bibnamefont {Ootuka}},\ }\href {https://doi.org/10.1143/jpsj.66.3722} {\bibfield  {journal} {\bibinfo  {journal} {J. Phys. Soc. Jpn.}\ }\textbf {\bibinfo {volume} {66}},\ \bibinfo {pages} {3722–3724} (\bibinfo {year} {1997})}\BibitemShut {NoStop}%
\bibitem [{\citenamefont {Penttil\"a}\ \emph {et~al.}(1999)\citenamefont {Penttil\"a}, \citenamefont {Parts}, \citenamefont {Hakonen}, \citenamefont {Paalanen},\ and\ \citenamefont {Sonin}}]{Exp_2}%
  \BibitemOpen
  \bibfield  {author} {\bibinfo {author} {\bibfnamefont {J.~S.}\ \bibnamefont {Penttil\"a}}, \bibinfo {author} {\bibfnamefont {U.}~\bibnamefont {Parts}}, \bibinfo {author} {\bibfnamefont {P.~J.}\ \bibnamefont {Hakonen}}, \bibinfo {author} {\bibfnamefont {M.~A.}\ \bibnamefont {Paalanen}},\ and\ \bibinfo {author} {\bibfnamefont {E.~B.}\ \bibnamefont {Sonin}},\ }\href {https://doi.org/10.1103/PhysRevLett.82.1004} {\bibfield  {journal} {\bibinfo  {journal} {Phys. Rev. Lett.}\ }\textbf {\bibinfo {volume} {82}},\ \bibinfo {pages} {1004} (\bibinfo {year} {1999})}\BibitemShut {NoStop}%
\bibitem [{\citenamefont {Kuzmin}\ \emph {et~al.}(1991)\citenamefont {Kuzmin}, \citenamefont {Nazarov}, \citenamefont {Haviland}, \citenamefont {Delsing},\ and\ \citenamefont {Claeson}}]{Exp_3}%
  \BibitemOpen
  \bibfield  {author} {\bibinfo {author} {\bibfnamefont {L.~S.}\ \bibnamefont {Kuzmin}}, \bibinfo {author} {\bibfnamefont {Y.~V.}\ \bibnamefont {Nazarov}}, \bibinfo {author} {\bibfnamefont {D.~B.}\ \bibnamefont {Haviland}}, \bibinfo {author} {\bibfnamefont {P.}~\bibnamefont {Delsing}},\ and\ \bibinfo {author} {\bibfnamefont {T.}~\bibnamefont {Claeson}},\ }\href {https://doi.org/10.1103/PhysRevLett.67.1161} {\bibfield  {journal} {\bibinfo  {journal} {Phys. Rev. Lett.}\ }\textbf {\bibinfo {volume} {67}},\ \bibinfo {pages} {1161} (\bibinfo {year} {1991})}\BibitemShut {NoStop}%
\bibitem [{\citenamefont {Josephson}(1962)}]{Josephson1}%
  \BibitemOpen
  \bibfield  {author} {\bibinfo {author} {\bibfnamefont {B.}~\bibnamefont {Josephson}},\ }\href {https://doi.org/10.1016/0031-9163(62)91369-0} {\bibfield  {journal} {\bibinfo  {journal} {Phys. Lett.}\ }\textbf {\bibinfo {volume} {1}},\ \bibinfo {pages} {251–253} (\bibinfo {year} {1962})}\BibitemShut {NoStop}%
\bibitem [{\citenamefont {Josephson}(1974)}]{Josephson2}%
  \BibitemOpen
  \bibfield  {author} {\bibinfo {author} {\bibfnamefont {B.~D.}\ \bibnamefont {Josephson}},\ }\href {https://doi.org/10.1103/RevModPhys.46.251} {\bibfield  {journal} {\bibinfo  {journal} {Rev. Mod. Phys.}\ }\textbf {\bibinfo {volume} {46}},\ \bibinfo {pages} {251} (\bibinfo {year} {1974})}\BibitemShut {NoStop}%
\bibitem [{\citenamefont {Likharev}\ and\ \citenamefont {Zorin}(1985)}]{Likharev}%
  \BibitemOpen
  \bibfield  {author} {\bibinfo {author} {\bibfnamefont {K.}~\bibnamefont {Likharev}}\ and\ \bibinfo {author} {\bibfnamefont {A.}~\bibnamefont {Zorin}},\ }\href {https://doi.org/https://doi.org/10.1007/BF00683782} {\bibfield  {journal} {\bibinfo  {journal} {J. Low. Temp. Phys.}\ }\textbf {\bibinfo {volume} {59}},\ \bibinfo {pages} {347–382} (\bibinfo {year} {1985})}\BibitemShut {NoStop}%
\bibitem [{\citenamefont {Ambegaokar}\ and\ \citenamefont {Baratoff}(1963)}]{Ambegaokar_Josephson_effect}%
  \BibitemOpen
  \bibfield  {author} {\bibinfo {author} {\bibfnamefont {V.}~\bibnamefont {Ambegaokar}}\ and\ \bibinfo {author} {\bibfnamefont {A.}~\bibnamefont {Baratoff}},\ }\href {https://doi.org/10.1103/PhysRevLett.10.486} {\bibfield  {journal} {\bibinfo  {journal} {Phys. Rev. Lett.}\ }\textbf {\bibinfo {volume} {10}},\ \bibinfo {pages} {486} (\bibinfo {year} {1963})}\BibitemShut {NoStop}%
\bibitem [{\citenamefont {Sch\"{o}n}\ and\ \citenamefont {Zaikin}(1990)}]{Schon_Zaikin_review}%
  \BibitemOpen
  \bibfield  {author} {\bibinfo {author} {\bibfnamefont {G.}~\bibnamefont {Sch\"{o}n}}\ and\ \bibinfo {author} {\bibfnamefont {A.}~\bibnamefont {Zaikin}},\ }\href {https://doi.org/10.1016/0370-1573(90)90156-v} {\bibfield  {journal} {\bibinfo  {journal} {Phys. Rep.}\ }\textbf {\bibinfo {volume} {198}},\ \bibinfo {pages} {237–412} (\bibinfo {year} {1990})}\BibitemShut {NoStop}%
\bibitem [{\citenamefont {Fazio}\ and\ \citenamefont {van~der Zant}(2001)}]{Fazio_review}%
  \BibitemOpen
  \bibfield  {author} {\bibinfo {author} {\bibfnamefont {R.}~\bibnamefont {Fazio}}\ and\ \bibinfo {author} {\bibfnamefont {H.}~\bibnamefont {van~der Zant}},\ }\href {https://doi.org/10.1016/s0370-1573(01)00022-9} {\bibfield  {journal} {\bibinfo  {journal} {Phys. Rep.}\ }\textbf {\bibinfo {volume} {355}},\ \bibinfo {pages} {235} (\bibinfo {year} {2001})}\BibitemShut {NoStop}%
\bibitem [{\citenamefont {Eckern}\ \emph {et~al.}(1984)\citenamefont {Eckern}, \citenamefont {Sch\"on},\ and\ \citenamefont {Ambegaokar}}]{PhysRevB_Schon}%
  \BibitemOpen
  \bibfield  {author} {\bibinfo {author} {\bibfnamefont {U.}~\bibnamefont {Eckern}}, \bibinfo {author} {\bibfnamefont {G.}~\bibnamefont {Sch\"on}},\ and\ \bibinfo {author} {\bibfnamefont {V.}~\bibnamefont {Ambegaokar}},\ }\href {https://doi.org/10.1103/PhysRevB.30.6419} {\bibfield  {journal} {\bibinfo  {journal} {Phys. Rev. B}\ }\textbf {\bibinfo {volume} {30}},\ \bibinfo {pages} {6419} (\bibinfo {year} {1984})}\BibitemShut {NoStop}%
\bibitem [{\citenamefont {Ambegaokar}\ \emph {et~al.}(1982)\citenamefont {Ambegaokar}, \citenamefont {Eckern},\ and\ \citenamefont {Sch\"on}}]{PhysRevLett_Schon}%
  \BibitemOpen
  \bibfield  {author} {\bibinfo {author} {\bibfnamefont {V.}~\bibnamefont {Ambegaokar}}, \bibinfo {author} {\bibfnamefont {U.}~\bibnamefont {Eckern}},\ and\ \bibinfo {author} {\bibfnamefont {G.}~\bibnamefont {Sch\"on}},\ }\href {https://doi.org/10.1103/PhysRevLett.48.1745} {\bibfield  {journal} {\bibinfo  {journal} {Phys. Rev. Lett.}\ }\textbf {\bibinfo {volume} {48}},\ \bibinfo {pages} {1745} (\bibinfo {year} {1982})}\BibitemShut {NoStop}%
\bibitem [{\citenamefont {Feynman}\ \emph {et~al.}(2011)\citenamefont {Feynman}, \citenamefont {Leighton},\ and\ \citenamefont {Sands}}]{Feynman3}%
  \BibitemOpen
  \bibfield  {author} {\bibinfo {author} {\bibfnamefont {R.~P.}\ \bibnamefont {Feynman}}, \bibinfo {author} {\bibfnamefont {R.~B.}\ \bibnamefont {Leighton}},\ and\ \bibinfo {author} {\bibfnamefont {M.}~\bibnamefont {Sands}},\ }\href@noop {} {\emph {\bibinfo {title} {The Feynman Lectures on Physics, Vol. III: The New Millennium Edition: Quantum Mechanics}}}\ (\bibinfo  {publisher} {Basic Books},\ \bibinfo {year} {2011})\BibitemShut {NoStop}%
\bibitem [{\citenamefont {Caldeira}\ and\ \citenamefont {Leggett}(1981)}]{PhysRevLettCL1}%
  \BibitemOpen
  \bibfield  {author} {\bibinfo {author} {\bibfnamefont {A.~O.}\ \bibnamefont {Caldeira}}\ and\ \bibinfo {author} {\bibfnamefont {A.~J.}\ \bibnamefont {Leggett}},\ }\href {https://doi.org/10.1103/PhysRevLett.46.211} {\bibfield  {journal} {\bibinfo  {journal} {Phys. Rev. Lett.}\ }\textbf {\bibinfo {volume} {46}},\ \bibinfo {pages} {211} (\bibinfo {year} {1981})}\BibitemShut {NoStop}%
\bibitem [{\citenamefont {Devoret}\ \emph {et~al.}(1995)\citenamefont {Devoret} \emph {et~al.}}]{devoret}%
  \BibitemOpen
  \bibfield  {author} {\bibinfo {author} {\bibfnamefont {M.~H.}\ \bibnamefont {Devoret}} \emph {et~al.},\ }\href@noop {} {\bibfield  {journal} {\bibinfo  {journal} {Les Houches, Session LXIII}\ }\textbf {\bibinfo {volume} {7}},\ \bibinfo {pages} {133} (\bibinfo {year} {1995})}\BibitemShut {NoStop}%
\bibitem [{\citenamefont {Fetter}\ and\ \citenamefont {Walecka}(2003)}]{Fetter}%
  \BibitemOpen
  \bibfield  {author} {\bibinfo {author} {\bibfnamefont {A.~L.}\ \bibnamefont {Fetter}}\ and\ \bibinfo {author} {\bibfnamefont {J.~D.}\ \bibnamefont {Walecka}},\ }\href@noop {} {\emph {\bibinfo {title} {Quantum Theory Of Many Particle Systems}}}\ (\bibinfo  {publisher} {Dover Publications},\ \bibinfo {year} {2003})\BibitemShut {NoStop}%
\bibitem [{\citenamefont {Mahan}(2000)}]{Mahan}%
  \BibitemOpen
  \bibfield  {author} {\bibinfo {author} {\bibfnamefont {G.~D.}\ \bibnamefont {Mahan}},\ }\href@noop {} {\emph {\bibinfo {title} {Many-Particle Physics}}}\ (\bibinfo  {publisher} {Springer},\ \bibinfo {year} {2000})\BibitemShut {NoStop}%
\bibitem [{\citenamefont {Shastry}(2006)}]{Shastry_PRB}%
  \BibitemOpen
  \bibfield  {author} {\bibinfo {author} {\bibfnamefont {B.~S.}\ \bibnamefont {Shastry}},\ }\href {https://doi.org/10.1103/PhysRevB.73.085117} {\bibfield  {journal} {\bibinfo  {journal} {Phys. Rev. B}\ }\textbf {\bibinfo {volume} {73}},\ \bibinfo {pages} {085117} (\bibinfo {year} {2006})}\BibitemShut {NoStop}%
\bibitem [{\citenamefont {Kirchner}\ \emph {et~al.}(1999)\citenamefont {Kirchner}, \citenamefont {Evertz},\ and\ \citenamefont {Hanke}}]{Evertz_PRB}%
  \BibitemOpen
  \bibfield  {author} {\bibinfo {author} {\bibfnamefont {S.}~\bibnamefont {Kirchner}}, \bibinfo {author} {\bibfnamefont {H.~G.}\ \bibnamefont {Evertz}},\ and\ \bibinfo {author} {\bibfnamefont {W.}~\bibnamefont {Hanke}},\ }\href {https://doi.org/10.1103/PhysRevB.59.1825} {\bibfield  {journal} {\bibinfo  {journal} {Phys. Rev. B}\ }\textbf {\bibinfo {volume} {59}},\ \bibinfo {pages} {1825} (\bibinfo {year} {1999})}\BibitemShut {NoStop}%
\bibitem [{\citenamefont {Jarrell}\ and\ \citenamefont {Gubernatis}(1996)}]{jarrell}%
  \BibitemOpen
  \bibfield  {author} {\bibinfo {author} {\bibfnamefont {M.}~\bibnamefont {Jarrell}}\ and\ \bibinfo {author} {\bibfnamefont {J.~E.}\ \bibnamefont {Gubernatis}},\ }\href {https://doi.org/10.1016/0370-1573(95)00074-7} {\bibfield  {journal} {\bibinfo  {journal} {Phys. Rep.}\ }\textbf {\bibinfo {volume} {269}},\ \bibinfo {pages} {133} (\bibinfo {year} {1996})}\BibitemShut {NoStop}%
\bibitem [{\citenamefont {Houzet}\ \emph {et~al.}(2024)\citenamefont {Houzet}, \citenamefont {Yamamoto},\ and\ \citenamefont {Glazman}}]{microwave}%
  \BibitemOpen
  \bibfield  {author} {\bibinfo {author} {\bibfnamefont {M.}~\bibnamefont {Houzet}}, \bibinfo {author} {\bibfnamefont {T.}~\bibnamefont {Yamamoto}},\ and\ \bibinfo {author} {\bibfnamefont {L.~I.}\ \bibnamefont {Glazman}},\ }\href {https://doi.org/10.1103/PhysRevB.109.155431} {\bibfield  {journal} {\bibinfo  {journal} {Phys. Rev. B}\ }\textbf {\bibinfo {volume} {109}},\ \bibinfo {pages} {155431} (\bibinfo {year} {2024})}\BibitemShut {NoStop}%
\bibitem [{\citenamefont {Minnhagen}(1985{\natexlab{a}})}]{Minn_PRB1}%
  \BibitemOpen
  \bibfield  {author} {\bibinfo {author} {\bibfnamefont {P.}~\bibnamefont {Minnhagen}},\ }\href {https://doi.org/10.1103/PhysRevB.32.3088} {\bibfield  {journal} {\bibinfo  {journal} {Phys. Rev. B}\ }\textbf {\bibinfo {volume} {32}},\ \bibinfo {pages} {3088} (\bibinfo {year} {1985}{\natexlab{a}})}\BibitemShut {NoStop}%
\bibitem [{\citenamefont {Weber}\ and\ \citenamefont {Minnhagen}(1988)}]{Minn_PRB2}%
  \BibitemOpen
  \bibfield  {author} {\bibinfo {author} {\bibfnamefont {H.}~\bibnamefont {Weber}}\ and\ \bibinfo {author} {\bibfnamefont {P.}~\bibnamefont {Minnhagen}},\ }\href {https://doi.org/10.1103/PhysRevB.37.5986} {\bibfield  {journal} {\bibinfo  {journal} {Phys. Rev. B}\ }\textbf {\bibinfo {volume} {37}},\ \bibinfo {pages} {5986} (\bibinfo {year} {1988})}\BibitemShut {NoStop}%
\bibitem [{\citenamefont {Minnhagen}(1985{\natexlab{b}})}]{Minn_PRL}%
  \BibitemOpen
  \bibfield  {author} {\bibinfo {author} {\bibfnamefont {P.}~\bibnamefont {Minnhagen}},\ }\href {https://doi.org/10.1103/PhysRevLett.54.2351} {\bibfield  {journal} {\bibinfo  {journal} {Phys. Rev. Lett.}\ }\textbf {\bibinfo {volume} {54}},\ \bibinfo {pages} {2351} (\bibinfo {year} {1985}{\natexlab{b}})}\BibitemShut {NoStop}%
\bibitem [{\citenamefont {{Korshunov}}(1987)}]{korshunov}%
  \BibitemOpen
  \bibfield  {author} {\bibinfo {author} {\bibfnamefont {S.~E.}\ \bibnamefont {{Korshunov}}},\ }\href@noop {} {\bibfield  {journal} {\bibinfo  {journal} {Pis'ma Zh. Eksp. Teor. Fiz}\ }\textbf {\bibinfo {volume} {45}},\ \bibinfo {pages} {342} (\bibinfo {year} {1987})}\BibitemShut {NoStop}%
\bibitem [{\citenamefont {Werner}\ and\ \citenamefont {Troyer}(2005{\natexlab{a}})}]{Troyer_MC}%
  \BibitemOpen
  \bibfield  {author} {\bibinfo {author} {\bibfnamefont {P.}~\bibnamefont {Werner}}\ and\ \bibinfo {author} {\bibfnamefont {M.}~\bibnamefont {Troyer}},\ }\href {https://doi.org/10.1143/PTPS.160.395} {\bibfield  {journal} {\bibinfo  {journal} {Prog. Theor. Phys. Supp.}\ }\textbf {\bibinfo {volume} {160}},\ \bibinfo {pages} {395} (\bibinfo {year} {2005}{\natexlab{a}})}\BibitemShut {NoStop}%
\bibitem [{\citenamefont {Herrero}\ and\ \citenamefont {Zaikin}(2002)}]{Herrero_Zaikin}%
  \BibitemOpen
  \bibfield  {author} {\bibinfo {author} {\bibfnamefont {C.~P.}\ \bibnamefont {Herrero}}\ and\ \bibinfo {author} {\bibfnamefont {A.~D.}\ \bibnamefont {Zaikin}},\ }\href {https://doi.org/10.1103/PhysRevB.65.104516} {\bibfield  {journal} {\bibinfo  {journal} {Phys. Rev. B}\ }\textbf {\bibinfo {volume} {65}},\ \bibinfo {pages} {104516} (\bibinfo {year} {2002})}\BibitemShut {NoStop}%
\bibitem [{\citenamefont {Remez}\ \emph {et~al.}(2024)\citenamefont {Remez}, \citenamefont {Kurilovich}, \citenamefont {Rieger},\ and\ \citenamefont {Glazman}}]{glazman1}%
  \BibitemOpen
  \bibfield  {author} {\bibinfo {author} {\bibfnamefont {B.}~\bibnamefont {Remez}}, \bibinfo {author} {\bibfnamefont {V.~D.}\ \bibnamefont {Kurilovich}}, \bibinfo {author} {\bibfnamefont {M.}~\bibnamefont {Rieger}},\ and\ \bibinfo {author} {\bibfnamefont {L.~I.}\ \bibnamefont {Glazman}},\ }\href {https://doi.org/10.1103/PhysRevB.110.054508} {\bibfield  {journal} {\bibinfo  {journal} {Phys. Rev. B}\ }\textbf {\bibinfo {volume} {110}},\ \bibinfo {pages} {054508} (\bibinfo {year} {2024})}\BibitemShut {NoStop}%
\bibitem [{\citenamefont {Houzet}\ and\ \citenamefont {Glazman}(2020)}]{glazman2}%
  \BibitemOpen
  \bibfield  {author} {\bibinfo {author} {\bibfnamefont {M.}~\bibnamefont {Houzet}}\ and\ \bibinfo {author} {\bibfnamefont {L.~I.}\ \bibnamefont {Glazman}},\ }\href {https://doi.org/10.1103/PhysRevLett.125.267701} {\bibfield  {journal} {\bibinfo  {journal} {Phys. Rev. Lett.}\ }\textbf {\bibinfo {volume} {125}},\ \bibinfo {pages} {267701} (\bibinfo {year} {2020})}\BibitemShut {NoStop}%
\bibitem [{\citenamefont {Kuzmin}\ \emph {et~al.}(2021)\citenamefont {Kuzmin}, \citenamefont {Grabon}, \citenamefont {Mehta}, \citenamefont {Burshtein}, \citenamefont {Goldstein}, \citenamefont {Houzet}, \citenamefont {Glazman},\ and\ \citenamefont {Manucharyan}}]{glazman3}%
  \BibitemOpen
  \bibfield  {author} {\bibinfo {author} {\bibfnamefont {R.}~\bibnamefont {Kuzmin}}, \bibinfo {author} {\bibfnamefont {N.}~\bibnamefont {Grabon}}, \bibinfo {author} {\bibfnamefont {N.}~\bibnamefont {Mehta}}, \bibinfo {author} {\bibfnamefont {A.}~\bibnamefont {Burshtein}}, \bibinfo {author} {\bibfnamefont {M.}~\bibnamefont {Goldstein}}, \bibinfo {author} {\bibfnamefont {M.}~\bibnamefont {Houzet}}, \bibinfo {author} {\bibfnamefont {L.~I.}\ \bibnamefont {Glazman}},\ and\ \bibinfo {author} {\bibfnamefont {V.~E.}\ \bibnamefont {Manucharyan}},\ }\href {https://doi.org/10.1103/PhysRevLett.126.197701} {\bibfield  {journal} {\bibinfo  {journal} {Phys. Rev. Lett.}\ }\textbf {\bibinfo {volume} {126}},\ \bibinfo {pages} {197701} (\bibinfo {year} {2021})}\BibitemShut {NoStop}%
\bibitem [{\citenamefont {Kurilovich}\ \emph {et~al.}(2025)\citenamefont {Kurilovich}, \citenamefont {Remez},\ and\ \citenamefont {Glazman}}]{glazman4}%
  \BibitemOpen
  \bibfield  {author} {\bibinfo {author} {\bibfnamefont {V.~D.}\ \bibnamefont {Kurilovich}}, \bibinfo {author} {\bibfnamefont {B.}~\bibnamefont {Remez}},\ and\ \bibinfo {author} {\bibfnamefont {L.~I.}\ \bibnamefont {Glazman}},\ }\href {https://doi.org/10.1038/s41467-025-56411-x} {\bibfield  {journal} {\bibinfo  {journal} {Nature Commun.}\ }\textbf {\bibinfo {volume} {16}},\ \bibinfo {pages} {1384} (\bibinfo {year} {2025})}\BibitemShut {NoStop}%
\bibitem [{\citenamefont {Mori}(1965)}]{Mori}%
  \BibitemOpen
  \bibfield  {author} {\bibinfo {author} {\bibfnamefont {H.}~\bibnamefont {Mori}},\ }\href {https://doi.org/10.1143/PTP.33.423} {\bibfield  {journal} {\bibinfo  {journal} {Prog. Theor. Phys.}\ }\textbf {\bibinfo {volume} {33}},\ \bibinfo {pages} {423} (\bibinfo {year} {1965})}\BibitemShut {NoStop}%
\bibitem [{\citenamefont {Werner}\ and\ \citenamefont {Troyer}(2005{\natexlab{b}})}]{Troyer_PRL}%
  \BibitemOpen
  \bibfield  {author} {\bibinfo {author} {\bibfnamefont {P.}~\bibnamefont {Werner}}\ and\ \bibinfo {author} {\bibfnamefont {M.}~\bibnamefont {Troyer}},\ }\href {https://doi.org/10.1103/PhysRevLett.95.060201} {\bibfield  {journal} {\bibinfo  {journal} {Phys. Rev. Lett.}\ }\textbf {\bibinfo {volume} {95}},\ \bibinfo {pages} {060201} (\bibinfo {year} {2005}{\natexlab{b}})}\BibitemShut {NoStop}%
\bibitem [{\citenamefont {Wolff}(1989)}]{Wolff_PRL}%
  \BibitemOpen
  \bibfield  {author} {\bibinfo {author} {\bibfnamefont {U.}~\bibnamefont {Wolff}},\ }\href {https://doi.org/10.1103/PhysRevLett.62.361} {\bibfield  {journal} {\bibinfo  {journal} {Phys. Rev. Lett.}\ }\textbf {\bibinfo {volume} {62}},\ \bibinfo {pages} {361} (\bibinfo {year} {1989})}\BibitemShut {NoStop}%
\bibitem [{\citenamefont {Luijten}\ and\ \citenamefont {Bl\"{o}te}(1995)}]{Luijten1}%
  \BibitemOpen
  \bibfield  {author} {\bibinfo {author} {\bibfnamefont {E.}~\bibnamefont {Luijten}}\ and\ \bibinfo {author} {\bibfnamefont {H.~W.}\ \bibnamefont {Bl\"{o}te}},\ }\href {https://doi.org/10.1142/S0129183195000265} {\bibfield  {journal} {\bibinfo  {journal} {Int. J. Mod. Phys. C}\ }\textbf {\bibinfo {volume} {06}},\ \bibinfo {pages} {359} (\bibinfo {year} {1995})}\BibitemShut {NoStop}%
\bibitem [{\citenamefont {Luijten}(2000)}]{Luijten2}%
  \BibitemOpen
  \bibfield  {author} {\bibinfo {author} {\bibfnamefont {E.}~\bibnamefont {Luijten}},\ }in\ \href@noop {} {\emph {\bibinfo {booktitle} {Computer Simulation Studies in Condensed-Matter Physics XII}}},\ \bibinfo {editor} {edited by\ \bibinfo {editor} {\bibfnamefont {D.~P.}\ \bibnamefont {Landau}}, \bibinfo {editor} {\bibfnamefont {S.~P.}\ \bibnamefont {Lewis}},\ and\ \bibinfo {editor} {\bibfnamefont {H.-B.}\ \bibnamefont {Sch{\"u}ttler}}}\ (\bibinfo  {publisher} {Springer Berlin Heidelberg},\ \bibinfo {address} {Berlin, Heidelberg},\ \bibinfo {year} {2000})\ pp.\ \bibinfo {pages} {86--99}\BibitemShut {NoStop}%
\end{thebibliography}%




%
%

\end{document}